\documentclass[sigconf, nonacm, balance, pdfa]{acmart}
\usepackage{flushend}
\usepackage{subcaption}
\usepackage{listings}
\usepackage{xcolor}
\usepackage{float}
\usepackage{tcolorbox}
\usepackage{fancybox}
\usepackage[a-2b]{pdfx}
\usepackage{url}

\definecolor{keywordcolor}{RGB}{0,0,180}
\definecolor{commentcolor}{RGB}{0,128,0}
\definecolor{stringcolor}{RGB}{163,21,21}
\definecolor{backgroundcolor}{RGB}{248,248,255}
\definecolor{numbercolor}{RGB}{128,128,128}

\lstdefinestyle{custom}{
    backgroundcolor=\color{backgroundcolor},   
    basicstyle=\ttfamily\small,                
    keywordstyle=\color{keywordcolor}\bfseries,
    commentstyle=\color{commentcolor}\itshape, 
    stringstyle=\color{stringcolor},           
    numberstyle=\tiny\color{numbercolor},      
    numbers=left,                              
    numbersep=5pt,                             
    tabsize=4,                                 
    showstringspaces=false,                    
    breaklines=true,                           
    rulecolor=\color{black},                   
    captionpos=b,                              
    escapeinside={\%*}{*)},                    
}

\newtcolorbox{grayblock}{
  colback=gray!15,    
  boxrule=0pt,        
  arc=6pt,            
  left=4pt, right=4pt,
  top=4pt, bottom=4pt
}

\newcommand{\eatdot}[1]{}

\newcommand\vldbdoi{10.14778/3785297.3785299}
\newcommand\vldbpages{549 - 562}
\newcommand\vldbvolume{19}
\newcommand\vldbissue{4}
\newcommand\vldbyear{2025}
\newcommand\vldbauthors{\authors}
\newcommand\vldbtitle{\shorttitle} 
\newcommand\vldbavailabilityurl{https://github.com/cacheMon/Lazy-Promotions}
\newcommand\vldbpagestyle{empty} 

\newcommand{\jason}[1]{\textcolor{orange}{jason: #1}}
\newcommand{\jdel}[1]{\textcolor{orange}{\sout{#1}}}

\newcommand{\frank}[1]{\textcolor{green}{frank: #1}}

\newcommand{\bobdel}[1]{\textcolor{purple}{\sout{#1}}}
\newcommand{\bob}[1]{\textcolor{purple}{bob: #1}}

\newcommand{\rashmi}[1]{{\color{magenta}[R says: #1]}}

\newcommand{\rrem}[1]{{\color{brown}[R removed: #1]}}
\newcommand{\todo}[1]{\textcolor{red}{TODO: #1}}

\renewcommand{\todo}[1]{}

\renewcommand{\jason}[1]{}
\renewcommand{\jdel}[1]{}

\renewcommand{\frank}[1]{}

\renewcommand{\bobdel}[1]{}
\renewcommand{\bob}[1]{}

\renewcommand{\rashmi}[1]{}

\renewcommand{\rrem}[1]{}

\newcommand{\boldparagraph}[1]{\noindent \textbf{#1}}

\usepackage{xspace}
\usepackage{cleveref}

\newcommand{\lp}{\textsc{Lazy Promotion} }

\newcommand{\delay}{Delay-LRU\xspace}
\newcommand{\delayparam}{\textit{delay ratio}\xspace}

\newcommand{\prob}{Prob\-a\-bi\-lis\-tic-LRU\xspace}
\newcommand{\probparam}{\textit{prob}\xspace}

\newcommand{\batch}{Batch-LRU\xspace}
\newcommand{\batchparam}{\textit{batch ratio}\xspace}
\newcommand{\batchparams}{\textit{batch ratios}\xspace}

\newcommand{\random}{Random-LRU\xspace}
\newcommand{\randomparam}{\textit{sample size}\xspace}

\newcommand{\clock}{FIFO-reinsertion\xspace}
\newcommand{\clockparam}{frequency bits\xspace}

\newcommand{\beladyfr}{offline FIFO-reinsertion\xspace}
\newcommand{\beladyrandom}{Belady-Random\xspace}
\newcommand{\beladyrandomLRU}{Belady-RandomLRU\xspace}

\newcommand{\offlinehotcache}{offline Hotcache\xspace}

\newcommand{\tofill}{{\color{red}{@@@}}\xspace}

\newcommand{\hotdelay}{Hot-delay\xspace}
\newcommand{\hotbatch}{Hot-batch\xspace}
\newcommand{\hotprob}{Hot-prob\xspace}
\newcommand{\hotcache}{Hotcache\xspace}
\newcommand{\hotcacheparam}{\textit{threshold}\xspace}

\newcommand{\dfr}{D-FR\xspace}
\newcommand{\age}{AGE\xspace}

\usepackage{enumitem}
\setlist{noitemsep, topsep=0pt}
\setlist{topsep=0pt, partopsep=0pt, parsep=0pt, itemsep=0pt}
\usepackage[font=small]{caption}

\begin{document}

\title{Demystifying and Improving Lazy Promotion in Cache Eviction}
\author{Qinghan Chen}
\affiliation{%
  \institution{Carnegie Mellon University}
  \city{Pittsburgh}
  \state{PA}
}
\email{qinghanc@andrew.cmu.edu}

\author{Muhammad Haekal Muhyidin Al-Araby}
\affiliation{
  \institution{Sepuluh Nopember Institute of Technology}
  \city{Surabaya}
  \country{Indonesia}
}
\email{5024221030@student.its.ac.id}

\author{Ziyue Qiu}
\affiliation{%
  \institution{Carnegie Mellon University}
  \city{Pittsburgh}
  \state{PA}
}
\email{ziyueqiu@andrew.cmu.edu}

\author{Zhuofan Chen}
\affiliation{%
  \institution{Carnegie Mellon University}
  \city{Pittsburgh}
  \state{PA}
}
\email{zhuofanc@andrew.cmu.edu}

\author{Rashmi Vinayak}
\affiliation{%
  \institution{Carnegie Mellon University}
  \city{Pittsburgh}
  \state{PA}
}
\email{rvinayak@cs.cmu.edu}

\author{Juncheng Yang}
\affiliation{%
\institution{Harvard University}
  \city{Cambridge}
  \state{MA}
}
\email{juncheng@seas.harvard.edu}


\begin{abstract}
Cache eviction algorithms play a critical role in the performance of modern data systems, yet their scalability is often limited by the high computational overhead associated with object promotions. \lp techniques have emerged as relaxations of traditional Least-Recently-Used (LRU) methods, designed to alleviate lock contention and increase throughput. This work uses production traces from real-world systems to benchmark five \lp strategies: \prob, \batch, \delay, \clock, and \random. We evaluate these techniques across miss ratio, scalability, promotion count, and a novel metric called promotion efficiency, which measures the number of hits per promotion.
Our results reveal that \delay and \clock significantly improve promotion efficiency, whereas \batch and \prob struggle to reduce promotions without significantly increasing miss ratio. We further explore the impact of lazy promotion in advanced algorithms such as ARC and 2Q and make a similar observation. 
Moreover, we uncover substantial optimization potential, showing that most cache promotions are unnecessary when equipped with oracle knowledge. To further reduce promotions in LRU, we propose two novel enhancements---Delayed \clock (D-FR) and Age-Guided Eviction (AGE)---that reduce promotions by 20–-60\% while achieving a similar or lower miss ratio. 
\end{abstract}

\maketitle

\pagestyle{\vldbpagestyle}
\begingroup\small\noindent\raggedright\textbf{PVLDB Reference Format:}\\
\vldbauthors. \vldbtitle. PVLDB, \vldbvolume(\vldbissue): \vldbpages, \vldbyear.\\
\href{https://doi.org/\vldbdoi}{doi:\vldbdoi}
\endgroup
\begingroup
\renewcommand\thefootnote{}\footnote{\noindent
This work is licensed under the Creative Commons BY-NC-ND 4.0 International License. Visit \url{https://creativecommons.org/licenses/by-nc-nd/4.0/} to view a copy of this license. For any use beyond those covered by this license, obtain permission by emailing \href{mailto:info@vldb.org}{info@vldb.org}. Copyright is held by the owner/author(s). Publication rights licensed to the VLDB Endowment. \\
\raggedright Proceedings of the VLDB Endowment, Vol. \vldbvolume, No. \vldbissue\ %
ISSN 2150-8097. \\
\href{https://doi.org/\vldbdoi}{doi:\vldbdoi} \\
}\addtocounter{footnote}{-1}\endgroup

\ifdefempty{\vldbavailabilityurl}{}{
\par\medskip
\begingroup\small\noindent\raggedright\textbf{PVLDB Artifact Availability:}\\
The source code and data are available at \url{\vldbavailabilityurl}.
\endgroup
}












\begin{figure}[ht]
    \centering
    \includegraphics[width = 0.45 \linewidth]{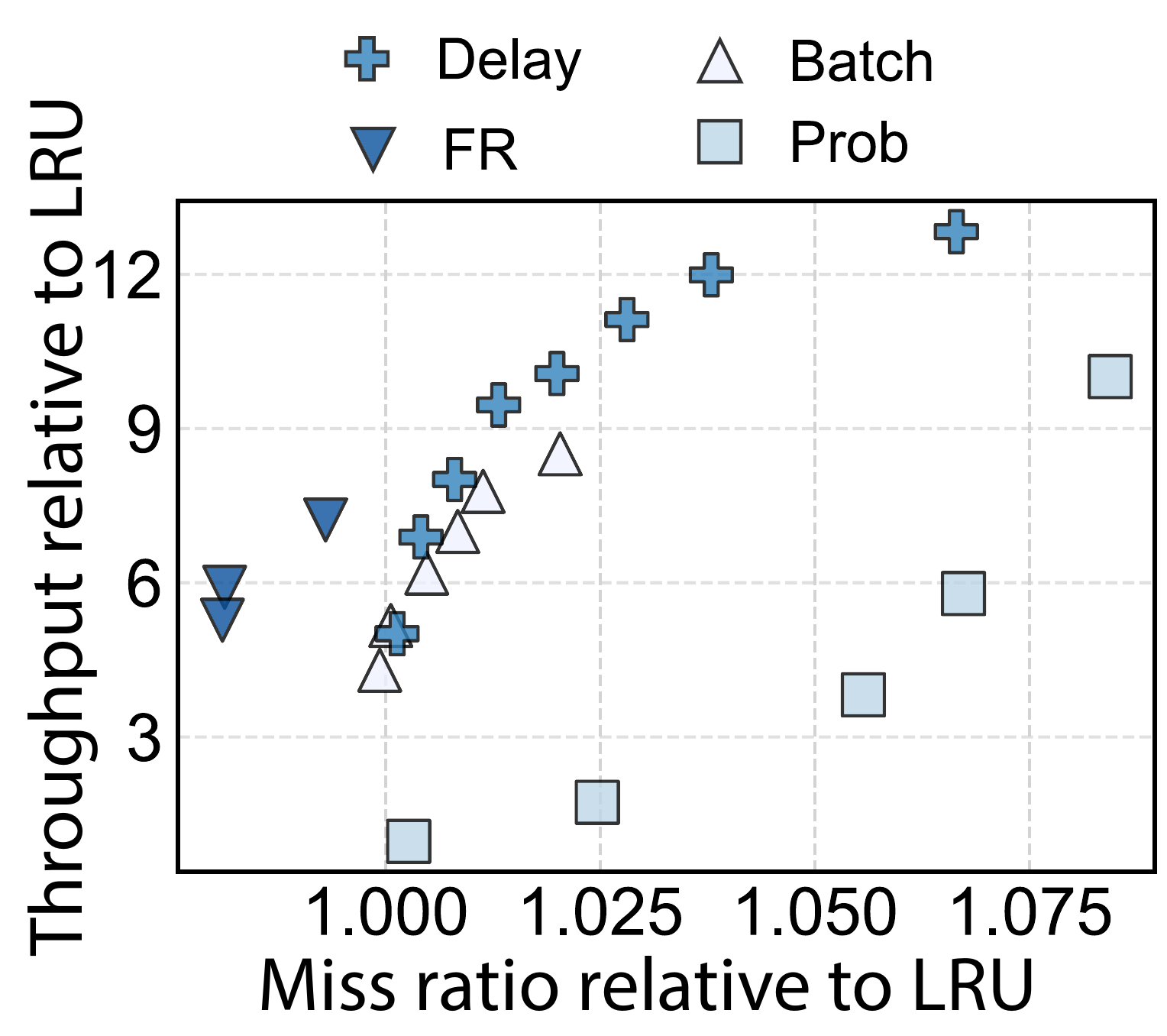}
    \includegraphics[width = 0.45 \linewidth, height=0.40 \linewidth]{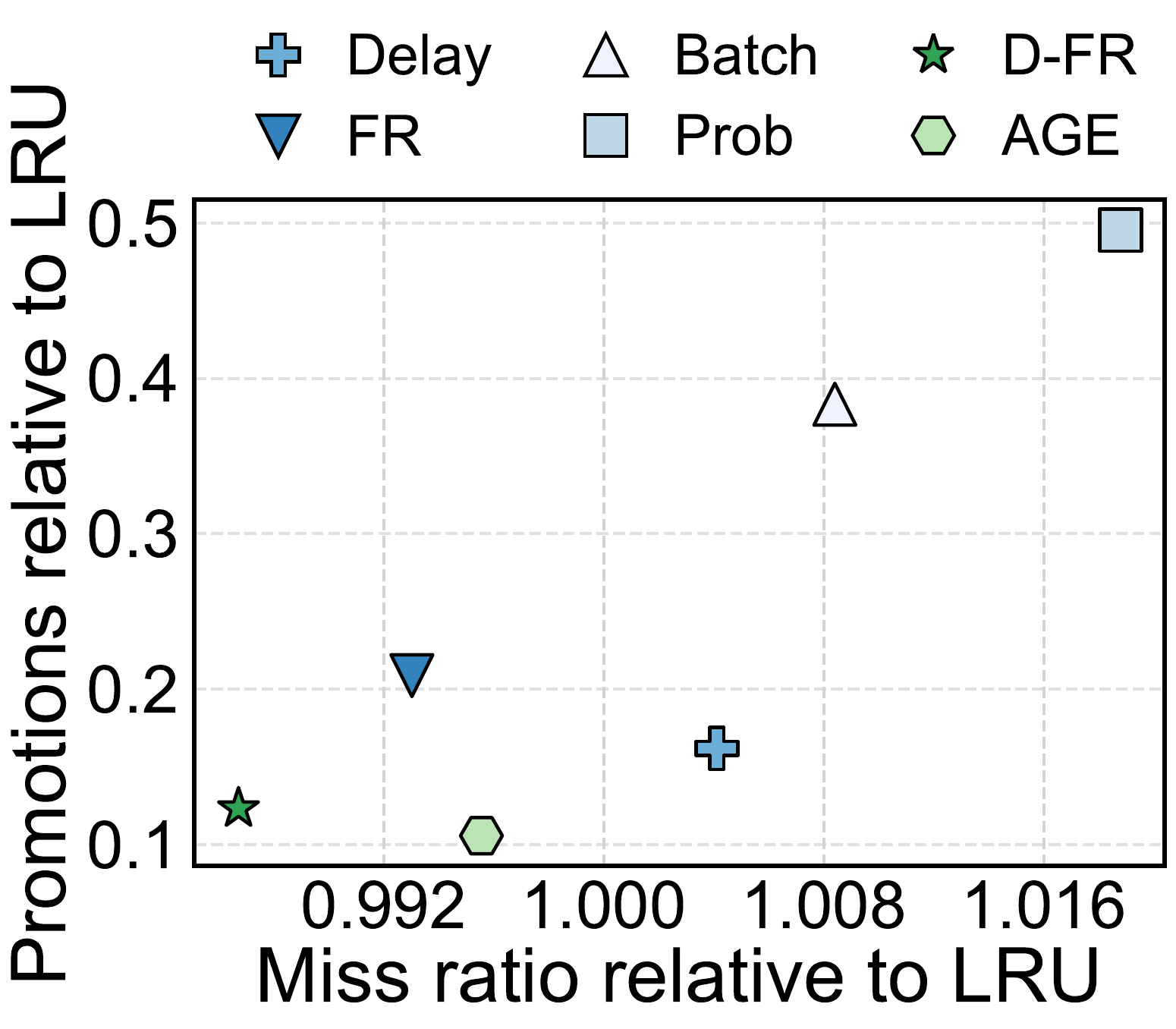}
    \Description{Overview and comparison of algorithm metrics, showing the effectiveness of different techniques.}
    \caption{Left: \normalfont{overview of \lp techniques. delay and \clock (FR) are more effective than \batch and \prob on reducing promotions and improving scalability (throughput at 16 threads) without significantly increasing the miss ratio.} \textbf{Right}: We introduce two simple techniques, \dfr and \age, to reduce promotions without increasing miss ratio.  }
    \label{fig:intro}
\end{figure}
\section{Introduction}

Caching plays an important role in data systems.
A cache stores a small portion of popular objects on a fast storage device, allowing requests to be served very quickly from the cache. Caching is now integral to nearly every layer of modern computer systems. 
For example, databases rely on the buffer pool to load data quickly~\cite{megiddo_arc_2003,jiang_lirs_2002}.
In MySQL, the InnoDB buffer pool \cite{kang2012innodb} acts as a cache, storing frequently accessed rows and index pages in memory. This reduces disk I/O, enabling faster data retrieval and improving query performance. 
Moreover, an operating system relies on the page cache to access data swiftly~\cite{linuxPageCache2Q}, and a content delivery network uses caching at the edge of the Internet to deliver images and videos to end users cheaply and quickly~\cite{mokhtarian2014caching}. 


The efficiency of a cache is measured by \textit{miss ratio}---the proportion of data requests that must be served by the backend. Reducing miss ratio allows more requests to be served from the cache and enables a faster and more responsive system. 
Besides efficiency, the other important metric of a cache is \textit{throughput}, which measures the hits a cache can serve every second.
A faster cache enables the operator to utilize fewer computational resources, thereby enhancing the utility of caching. Lastly, modern computer servers often have tens to hundreds of cores per socket~\cite{IBM-power7,chrysos2014xeonphi,amd2022genoa,ampere2020altra}, making \textit{scalability}---the capability to scale throughput with the number of cores, a critical requirement. 

At the core of caching is the cache eviction algorithm, which determines the objects stored in the cache and the order in which they are evicted. 
A good eviction algorithm maximizes the likelihood of cache hits, thus minimizing backend accesses and reducing data access latency. 
Least-Recently-Used (LRU) is the most popular eviction algorithm used in many production systems~\cite{cachelib, Varnish, redis}. 
LRU uses a doubly-linked list to maintain objects in the order of their last access time, positioning the most recently used item at the head and the least recently used at the tail. When an object is accessed, it is \textit{promoted} to the head of the list. During eviction, the object at the tail is removed. 
Because operations on items in the middle of a linked list cannot be performed atomically, all the operations in LRU require locking. This significantly limits the throughput of an LRU cache when using multiple CPU cores, as it must wait to acquire the lock. 
Although FIFO is more scalable than LRU, it has not been widely adopted because LRU is widely considered more efficient than FIFO, achieving a lower miss ratio~\cite{belady_study_1966, denning_working_2021}. 

As scalability has increasingly become a concern for modern cache systems, production system engineers have developed different relaxations of LRU promotions to mitigate the scalability bottleneck. For example, Meta Cachelib~\cite{berg_cachelib_2020} delays the promotion of a recently promoted object to reduce lock contention; Meta HHVM~\cite{ottoni2018hhvm} employs a probabilistic LRU using try-lock --- promote an object to the head of the linked list only upon a successful lock operation; Twitter Segcache~\cite{yang_segcache_2021} and Google Cliquemap~\cite{singhvi_cliquemap_2021} both use a batched eviction; and RocksDB~\cite{dong2021rocksdb} and PostgreSQL~\cite{postgresql} use \clock
to reduce lock contention.
Because these techniques primarily improve scalability by reducing the number of promotions and lock contention, similar to previous work~\cite{yang_fifo_2023_manual}, we refer to them as \lp techniques.

Despite the broad deployment of \lp techniques in production systems, there has not been a good understanding in the effectiveness of these techniques. 
It is often believed that \lp makes a tradeoff between cache efficiency and scalability. As a result, these relaxations are sometimes called weak LRU~\cite{jiang2005making, de2006bounded} because they sacrifice miss ratio to improve throughput and scalability. 

We implement and benchmark \lp techniques at scale to understand their effectiveness in improving scalability, reducing promotions, and maintaining miss ratio. We evaluate five different \lp techniques from real-world systems: \prob (\S\ref{sec:prob}), \batch (\S\ref{sec:batch}), \delay (\S\ref{sec:delay}), \clock (\S\ref{sec:clock}), and \random (\S\ref{sec:random}) using traces from 9 different companies (\Cref{table:workload}), including Alibaba~\cite{alibaba-trace1, alibaba-trace2}, Cloudphysics~\cite{waldspurger_efficient_2015}, Meta~\cite{meta-trace}, Microsoft~\cite{narayanan_write_2008}, Tencent\cite{tencentPhotoTrace}, Twitter~\cite{yang_large_2020}, Wikipedia~\cite{wikitrace}, and two content delivery networks. In total, we use 6357 traces of 346 billion requests for 2,818 TB of data.

We propose a new metric, \textit{promotion efficiency}, to evaluate the effectiveness of different \lp techniques. It measures the (average) number of hits a promotion leads to. A \lp technique with high promotion efficiency indicates that its promotions are more valuable. LRU has a mean promotion efficiency of 0.037---each promotion only leads to 0.037 hits. 
Among the \lp techniques, we find that \delay and \clock are most effective. When the miss ratio is within 1\% of LRU's performance, \delay and \clock have promotion efficiency of as high as 0.4 and 0.3 hits per promotion, respectively. \batch is worse, achieving less than 0.1 hits per promotion. 
\prob performs the worst because it randomly reduces promotion without distinguishing between popular and unpopular objects. 

Many advanced eviction algorithms consist of one or more LRU queues. Our findings are not limited to LRU. We add \lp to ARC~\cite{megiddo_arc_2003} and 2Q~\cite{johnson_2q_1994}. We find that with \delay and \clock can effectively reduce the number of promotions in ARC and 2Q, meanwhile \prob and \batch are less effective, often leading to significant increases in miss ratio. 

Using Belady's MIN~\cite{belady_study_1966} as a reference for optimal cache eviction, we find that most cache promotions in LRU are unnecessary, with only 10\% needed on average. This finding underscores the potential for optimization. Moreover, we find that promotion (reinsertion) on eviction (as in \clock) is strictly better than promotion on cache hits. If we equip \clock with future knowledge, it can reduce promotions by over 90\% on average across all 6,357 traces while also slightly reducing miss ratio. 


Leveraging our findings, we introduce two simple \lp techniques, Delayed \clock (D-FR) and Age-Guided Eviction (AGE), to further reduce the number of promotions and improve promotion efficiency. D-FR leverages delay in \clock to reduce promotions, while \age utilizes recency information to filter out unnecessary reinsertions in \clock. Our evaluations on production traces show that these techniques achieve a lower or similar miss ratio while further reducing promotions by $20\%-60\%$ and improving the promotion efficiency on average by more than 80\%. 

We make the following contributions in this paper:
\begin{itemize}[nosep, topsep=0pt, partopsep=0pt, parsep=0pt, leftmargin=*]
    \item We implemented and benchmarked the effectiveness of five different \lp techniques from real-world systems and integrated them into advanced algorithms, including ARC and 2Q, using 6357 production traces. 
    \item We introduced a new metric, promotion efficiency, to evaluate how well algorithms minimize promotions while keeping a low miss ratio. Our results show that \delay and \clock improve promotion efficiency. 
    \item We discovered that most promotions are unnecessary, suggesting a huge potential to further reduce promotions. 
    \item We present two simple techniques, \dfr and \age, to reduce the unnecessary promotions while maintaining a similar or lower miss ratio. Evaluation on production traces shows that they can reduce promotions by 20--60\% without impacting miss ratio.
\end{itemize}

\begin{table*}[t]
\small
\centering
\caption{Descriptions of different \lp techniques.}
\vspace{-1.0em}
\label{table:existing_lp}
\begin{tabular}{c|c|c|c}
\hline
Algorithm name & Algorithm parameter & Description & Deployed systems\\
\hline
\delay & \delayparam & Skip recent promotions & Meta Cachelib~\cite{cachelib} \\
\prob & \probparam & Skip promotions by chance & Meta HHVM~\cite{ottoni2018hhvm}\\
\batch & \batchparam & Batch promotions between fixed intervals & CliqueMap~\cite{singhvi_cliquemap_2021}, Ristretto~\cite{ristretto}  \\
\clock & \clockparam & Defers promotions to eviction & RocksDB~\cite{dong2021rocksdb}\\
\random & \randomparam & Sample eviction candidates and evict based on last access time & Redis~\cite{pan_predis_2019}\\
\hline
\end{tabular}
\end{table*}



\section{Background and motivation}\label{sec:bg}

\subsection{Software cache and eviction algorithm}\label{sec:bg:cache}

Caching plays a vital role in modern data systems, enabling efficient data access across various scenarios. In compute-storage disaggregated architectures, caching mitigates the high latency of network-based data transfers~\cite{wang2023disaggregated, zhang2020understanding, ruan2023persistent}. Similarly, database systems use caching to store frequently accessed data in memory, reducing disk I/O and improving query response times. 


At the core of effective caching lies the eviction algorithm, which determines which objects remain in a cache when it reaches capacity. Two of the most popular algorithms, Least Recently Used (LRU) and First-In-First-Out (FIFO), represent two extremes in managing cache hits, with respect to two critical metrics: \textit{efficiency} (measured by the miss ratio) and \textit{scalability} (measured by throughput, or the rate of requests served per second).

LRU and its variants are the most widely used algorithms in production~\cite{memcached, cachelib, linuxPageCache2Q, mySQLBuffer2Q, Nginx, Varnish, redis}. LRU retains the most recently accessed objects, based on the principle that recently accessed pages are likely to be reused in the near future~\cite{belady_study_1966, denning_working_2021}. LRU usually achieves high efficiency but poor scalability due to its need to \textit{promotion}: moving the accessed object to the head in a doubly linked list 
, a process that must be protected by a global lock to ensure consistency~\cite{yang_fifo_2023_manual, qiu_frozenhot_2023}. This lock contention restricts throughput growth as the number of CPU cores increases, making LRU poorly suited for highly parallel environments, such as databases. 

FIFO evicts the oldest objects without considering access patterns. It excels in scalability without promotions or locking during cache hits. However, its simplicity often results in poor efficiency, as it fails to retain frequently accessed data. 
The two algorithms represent opposite ends of the spectrum: LRU promotes every object upon a hit, while FIFO promotes none.



\subsection{Lazy promotion techniques}\label{sec:bg:lp}

To reduce the computation per cache hit and make LRU more scalable, \lp techniques have emerged in real-world systems as relaxations of the LRU algorithm~\cite{yang_fifo_2023_manual}. These techniques focus on reducing the frequency of promotions during cache hits while maintaining low miss ratios. In this subsection, we describe these \lp techniques, highlighting their underlying intuitions and design principles with a summary in~\Cref{table:existing_lp}.

\boldparagraph{\prob. }
A straightforward solution to reducing the number of promotions is to avoid them when they become a bottleneck. 
HHVM~\cite{ottoni2018hhvm}, developed by Meta, employs probabilistic promotion using try-lock. Promotions are randomly skipped based on lock availability: an object is promoted only when a thread successfully acquires a lock using try-lock. If the lock is unavailable, the promotion is bypassed. As a result, the likelihood of promotion is directly tied to the thread's ability to acquire the lock. Under high contention, most promotions are bypassed to avoid waiting for the lock, improving the scalability. 

\boldparagraph{\batch. }
Because each promotion requires an expensive lock operation, another solution to improve scalability is to increase the critical section size and reduce the number of locking operations. For example, BP-wrapper~\cite{ding_bp-wrapper_2009} batches the LRU promotions to improve scalability. Similarly, Google's CliqueMap~\cite{singhvi_cliquemap_2021} is a distributed caching system that also uses \batch for eviction. In distributed deployments, the cache resides on the server, and clients use one-sided Remote Direct Memory Access (RDMA) to retrieve data directly without communicating with the server. As a result, the server remains unaware of the access pattern and cannot decide which objects to evict when writing new objects. To address this problem, clients \textit{periodically} share requested objects with the cache server using RPC. This allows the server to perform batched promotions. 
This batching strategy reduces the promotion overhead in distributed caches and the scalability bottleneck in local caches. 

\boldparagraph{\delay. }
Most of the cache requests are for popular objects, which do not require frequent promotion to stay in the cache. Therefore, we do not need to promote an object if it was promoted recently. Moreover, it takes a long time for a promoted object to traverse through the cache before being evicted. As long as the next promotion happens before eviction, it is sufficient. Leveraging this insight, Meta CacheLib~\cite{berg_cachelib_2020} uses \delay. It employs a tunable parameter that controls the \delayparam, ensuring that promotions are triggered only after a specified duration has elapsed since the last promotion. Increasing the \delayparam reduces the number of promotions, improving scalability; however, doing so also takes the risk that useful objects might be evicted, increasing the miss ratio. 

\boldparagraph{\clock. } 
A traditional solution to reduce promotion overhead is to delay promotion until eviction time. The corresponding algorithm has different names based on the underlying implementation, e.g., CLOCK, FIFO-Reinsertion, or Second Chance.  
In the following text, we use \clock to refer to this algorithm. 
Because \clock does not need to move the requested object to the head upon each request, it can be implemented using a ring buffer or atomic updates on the linked list. This not only improves throughput but also scalability. For example, the new eviction algorithm for the block cache in  RocksDB~\cite{dong2021rocksdb} and PostgreSQL~\cite{postgresql} uses \clock to improve scalability

\boldparagraph{\random. }
Although a linked list is the most common data structure for an LRU cache implementation, it brings the unavoidable scalability problem. An alternative approach to implement an approximate LRU can leverage random sampling at eviction. For example, Redis~\cite{pan_predis_2019} employs a \random policy that randomly samples $k$ (e.g., 5) objects from the cache and evicts the least recently used one among them. This approach addresses the scalability issue of LRU by reducing the need for locking during cache hits. \random does not need to maintain a fully ordered list. Instead, each object tracks the last access time, updating which does not require locking, thus improving the thread scalability. 
Although \random is not a \lp technique technically because it does not perform any promotion, we include it in our study because it is also an approximate LRU that improves scalability.

\section{How do existing \lp techniques perform?} 
\label{sec:experiment}
This section presents our benchmarks and measurements of different \lp techniques.

\begin{table*}[t]
\small
\centering
\caption{Datasets used in this work. For old datasets, we exclude traces with fewer than 1 million requests.}
\label{table:workload}
\begin{tabular}{@{}lrrrrrrrrccc@{}}
\toprule
\multicolumn{1}{l}{Trace} & 
\multicolumn{1}{r}{Approx} & 
\multicolumn{1}{r}{Cache} & 
\multicolumn{1}{r}{time span} & 
\multicolumn{1}{r}{\# Traces} & 
\multicolumn{1}{r}{\# Request} & 
\multicolumn{1}{r}{Request} & 
\multicolumn{1}{r}{\# Object} & 
\multicolumn{1}{r}{Object} & \\
\multicolumn{1}{l}{collections} & \multicolumn{1}{r}{time} & \multicolumn{1}{r}{type} &\multicolumn{1}{r}{(days)} & \multicolumn{1}{r}{} & \multicolumn{1}{r}{(million)} & \multicolumn{1}{r}{(TB)} & \multicolumn{1}{r}{(million)} & \multicolumn{1}{r}{(TB)} \\
\midrule
MSR~\cite{narayanan_write_2008} & 2007 & Block & 7 & 14 & 410 & 9.8 & 73 & 3.0\\
FIU~\cite{fiutrace} & 2008--11 & Block & 9--28 & 9 & 513 & 1.6 & 19 & 0.056\\
Cloudphysics~\cite{waldspurger_efficient_2015} & 2015 & Block & 7 & 106 & 2,115 & 80.7 & 492 & 21.9\\
CDN 1 & 2018 & Object & 7 & 301 & 3,429 & 3,227 & 237 & 207\\
Tencent Photo~\cite{tencentPhotoTrace} & 2018 & Object & 8 & 2 & 5,649 & 147 & 1,038 & 23.7\\
WikiMedia CDN~\cite{wikitrace} & 2019 & Object & 7 & 4 & 12,400 & 195 & 249 & 13.1\\
Tencent CBS~\cite{tencentBlockTrace} & 2020 & Block & 8 & 4,048 & 33,268 & 1,143 & 2,113 & 108\\
Alibaba~\cite{alibaba-trace1, alibaba-trace2, alibaba-trace3} & 2020 & Block & 30 & 609 & 20,038 & 647 & 1,676 & 112\\
Twitter~\cite{yang_large_2020} & 2020 & KV & 7 & 51 & 233,965 & 103 & 17,342 & 4.9\\
CDN 2 & 2021 & Object & 7 & 1,205 & 20,470 & 2,079 & 1,232 & 732\\
Meta KV~\cite{meta-trace} & 2022 & KV & 1 & 5 & 13,314 & 8.2 & 371 & 0.43\\
Meta CDN~\cite{meta-trace} & 2023 & Object & 7 & 3 & 231 & 8,594 & 76 & 1,527\\ 
\bottomrule
\end{tabular}
\end{table*}

\boldparagraph{Dataset and testbed. } We used a diverse set of real-world traces from sources including Alibaba\cite{alibaba_block}, CloudPhysics\cite{waldspurger_efficient_2015}, FIU\cite{fiutrace}, Meta, Tencent\cite{tencent_block, tencent_photo}, Wikipedia\cite{wikitrace}, and two private CDN service companies for our measurements. These traces were collected between 2007 and 2023, covering key-value, block, and object caches. Altogether, the datasets encompass 6357 traces with \textit{346 billion requests} for \textit{25 billion objects}, with \textit{16,625 TB of traffic} for a total of \textit{2,818 TB of data}. More details can be found in \Cref{table:workload}.

\boldparagraph{Miss Ratio Measurement. }
We implemented each \lp technique on top of libCacheSim~\cite{libCacheSim_github} and replayed the traces in our dataset to measure miss ratio. 
We evaluate each technique using three cache sizes: 0.1\%, 1\%, and 10\% of the working set size (the total number of objects in the trace). Because the diverse traces exhibit different access patterns, the miss ratios at the same cache size, e.g., 1\% of the working set size, can range from less than 1\% to more than 80\%. 
We use the LRU miss ratio as a baseline, and calculate the relative miss ratio as \[ \textit{miss}_{\textit{A}} / \textit{miss}_{\textit{LRU}}\] for a technique $A$ being studied. 
We present the results using boxplots, where the box represents the 25th and 75th percentiles, and the whiskers represent the 10th and 90th percentiles. 
We used a cluster of c8220 nodes on Cloudlab for miss ratio measurements~\footnote{As the miss ratio is deterministic and independent of the underlying hardware, the specific hardware used does not affect the outcome of these measurements.}.

\begin{figure*}[ht]
    \centering
    \begin{subfigure}[b]{0.28\linewidth}
        \centering
        \includegraphics[width=\linewidth]{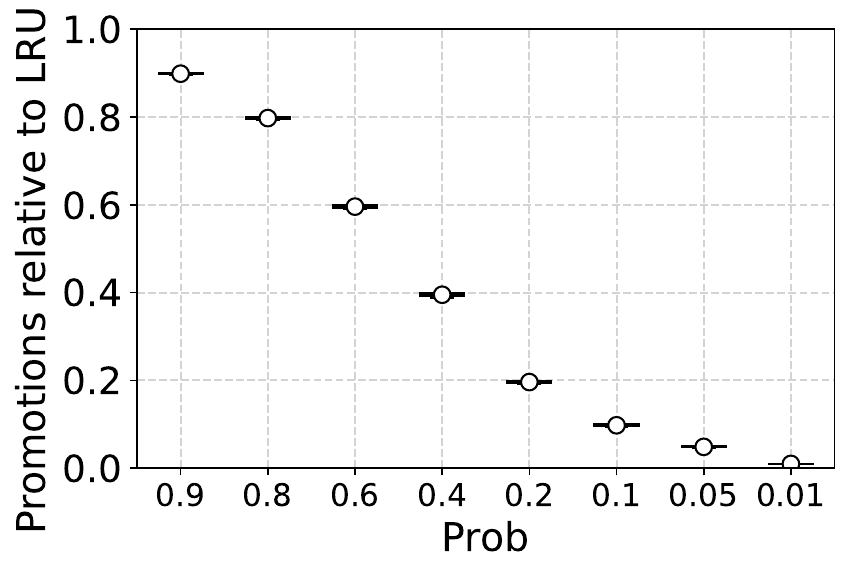}
        \caption{Relative number of promotions}
        \label{fig:prob_promotions}
    \end{subfigure}
   \hspace{0.8em}
    \begin{subfigure}[b]{0.28\linewidth}
        \centering
        \includegraphics[width=\linewidth]{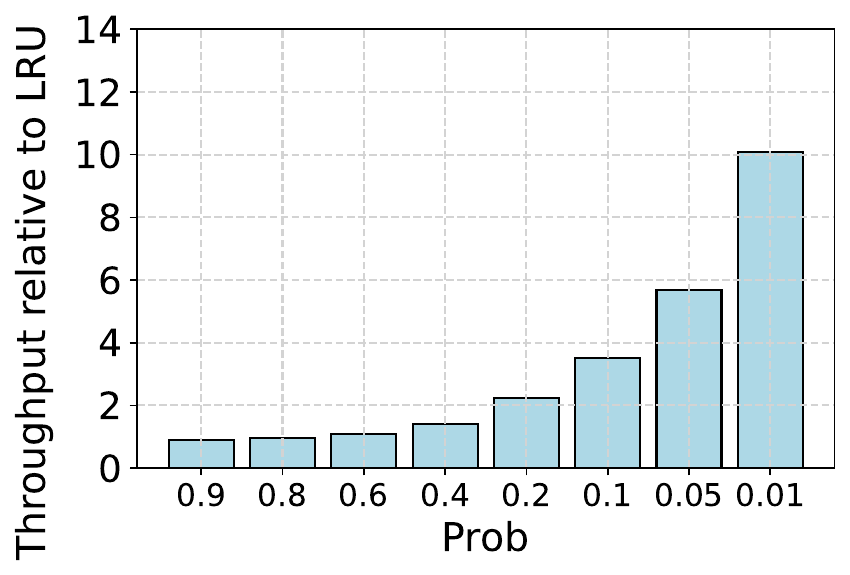}
        \caption{Throughput Imprv (16 threads)}
        \label{fig:prob_speedup}
    \end{subfigure}
   \hspace{0.8em}
    \begin{subfigure}[b]{0.29\linewidth}
        \centering
        \includegraphics[width=\linewidth]{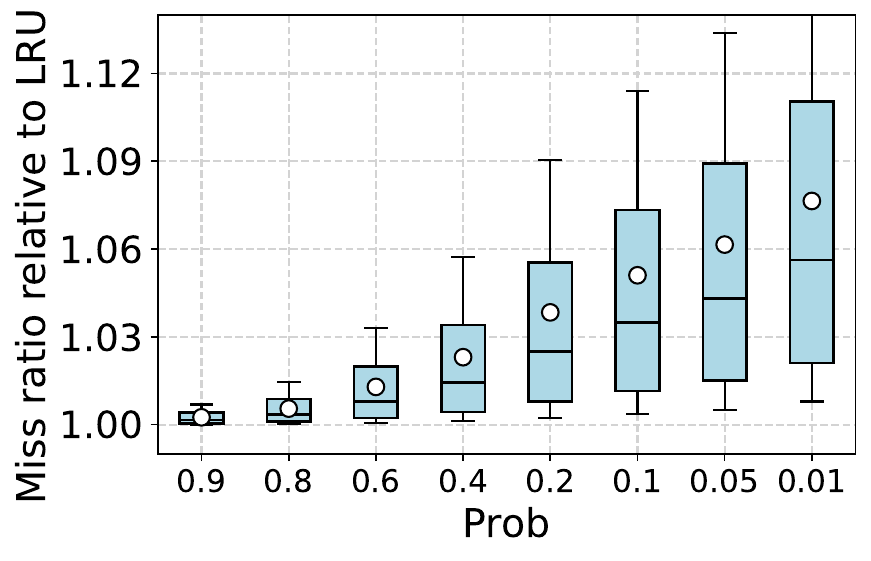}
        \caption{Relative miss ratio}
        \label{fig:prob_miss_ratio}
    \end{subfigure}
    \caption{\prob skips promotions randomly based on a given parameter \probparam. \normalfont{It reduces the number of promotions proportionally to the \probparam. However, the throughput does not increase significantly until the probability is reduced to very low. Meanwhile, \prob increases miss ratio even when the \prob is low.} }
    \label{fig:prob_miss_speedup_promotion}
\end{figure*}

\boldparagraph{Scalability Measurement. }
We implemented concurrent versions of each \lp technique. Unlike miss ratio measurements, which can be performed in parallel, the throughput/scalability measurements cannot be performed in parallel due to interference. Moreover, running all 6,357 real-world traces for every \lp technique and parameter configuration would require over a year to complete. Therefore, we performed the scalability measurement using a synthetic Zipfian trace consisting of 10 million requests for 1 million unique objects, with a skewness parameter of 1.0. The Zipfian parameter was selected to capture the cache access patterns observed in real-world workloads~\cite{atikoglu_workload_2012, berg_cachelib_2020, chen_hotring_2020, yang_large-scale_2021}. Because miss ratio can affect throughput, we chose different cache sizes for different \lp techniques so that they all achieve the same miss ratio. 
We experimented with both 10\% and 1\% miss ratios; however, we only present the 1\% miss ratio results due to space constraints. For the ease of visualization, we normalized the throughput of each \lp technique relative to the LRU baseline by calculating the ratio \[ throughput_A/throughput_{LRU}\] for a \lp technique $A$.
All measurements were conducted using r650 servers from Cloudlab, which are equipped with dual Intel Xeon Platinum 8360Y 36-core processors and 256GB of DDR4 DRAM. We disabled hyperthreading and turbo-boost, and limited our benchmark to use only one NUMA domain to ensure consistency across different benchmarks. Moreover, we repeated each experiment five times using randomly generated Zipfian traces and reported the mean results. 

\begin{figure*}[ht]
    \centering
    \begin{subfigure}[b]{0.28\linewidth}
        \centering
        \includegraphics[width=\linewidth]{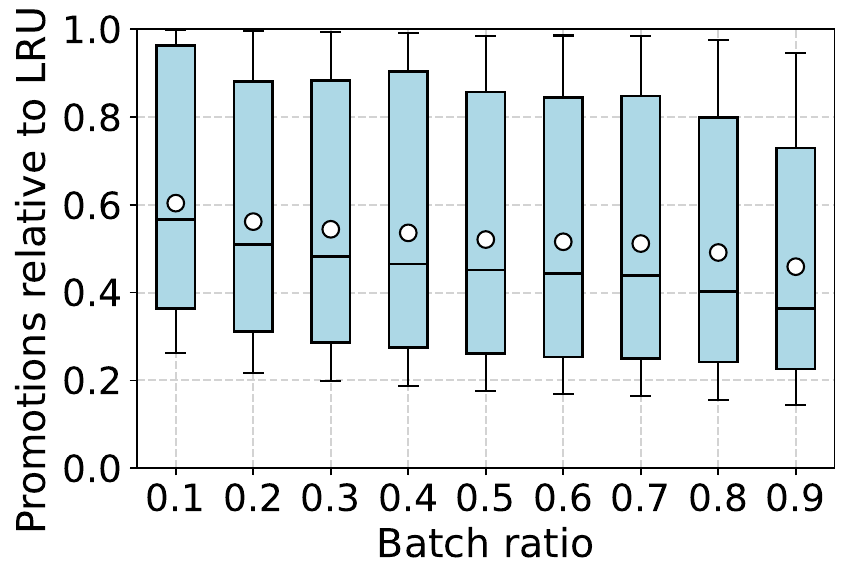}
        \caption{Relative number of promotions}
        \label{fig:batch_promotions}
    \end{subfigure}
   \hspace{0.8em}
    \begin{subfigure}[b]{0.28\linewidth}
        \centering
        \includegraphics[width=\linewidth]{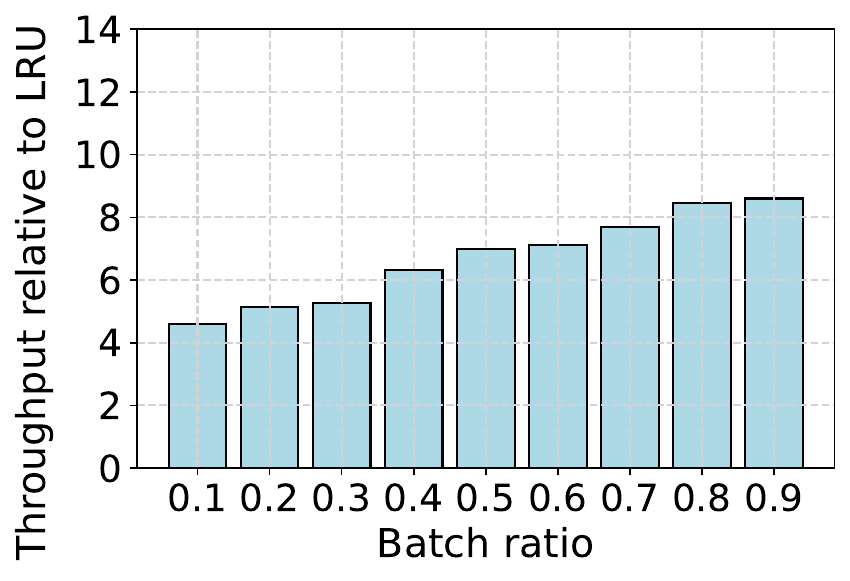}
        \caption{Throughput Imprv (16 threads)}
        \label{fig:batch_speedup}
    \end{subfigure}
   \hspace{0.8em}
    \begin{subfigure}[b]{0.29\linewidth}
        \centering
        \includegraphics[width=\linewidth]{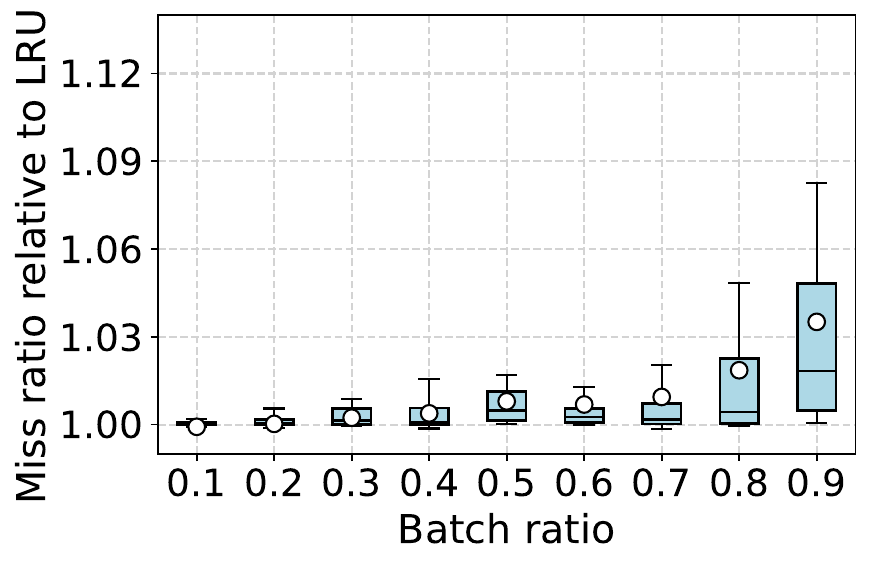}
        \caption{Relative miss ratio}
        \label{fig:batch_miss_ratio}
    \end{subfigure}
    \caption{\batch promotes objects in batches to reduce the number of promotions. \normalfont{A larger batch (\batchparam) leads to fewer promotions and higher throughput, but also a higher miss ratio.} }
    \label{fig:batch_miss_speedup_promotion}
\end{figure*}

\boldparagraph{Promotion Measurement. } While the throughput under different numbers of threads shows scalability, the throughput results depend on hardware and implementations. Therefore, we also measured the \textit{number of promotions} as a deterministic cost metric to capture the benefit of \lp. Recall that each promotion requires locking, which limits the scalability. Fewer promotions lead to fewer lock operations and better scalability. 
This metric is also computed relatively to LRU, as 
\[\textit{\# promos}_{\textit{A}} / \textit{\# promos}_{\textit{LRU}}\] 
for a \lp technique A.

\boldparagraph{Promotion Efficiency Measurement. }
i.e., \[(\textit{\#miss}_\textit{FIFO} - \textit{\#miss}_\textit{A})/\textit{\#promos}_\textit{A}\] 
We use FIFO as a baseline in this metric because it performs no promotion and is the extreme of a \lp technique. 
A high promotion efficiency indicates that the promotions in the technique are more valuable. 

Although we performed measurements using three different cache sizes —0.1\%, 1\%, and 10\% of the working set size —we found that the observations across different cache sizes are similar. Therefore, we only present the results at the 1\% cache size for space reasons. We will include the results of different cache sizes in our open-source repository. 

\subsection{\prob} \label{sec:prob}
\begin{lstlisting}[caption={promotion during cache hit in \prob}, style=custom, language=Python]
p = random(0,1)
if p < prob:
    promote(obj)
\end{lstlisting}
The root cause of LRU's poor scalability is the large number of promotions and corresponding locking operations. Therefore, \prob skips promotions randomly based on a given probability \probparam, which ranges from 0 to 1. 
When \probparam is 1, \prob becomes LRU. When \probparam is 0, \prob becomes FIFO~\footnote{Although we describe \probparam as a constant, \prob implementations in production systems are often more complex, and \probparam may change based on contention~\cite{trylock} as described in \Cref{sec:bg:lp}. }.

By randomly skipping promotions, \prob is very effective at reducing the number of promotions. \Cref{fig:prob_promotions} shows that the number of promotions decreases proportionally with the decrease of \probparam. For example, a \probparam of 0.1 can reduce the number of promotions by 90\%. 

Although \prob can significantly reduce the number of promotions, we observe that the throughput does not increase significantly when the probability is larger than 0.4 (\Cref{fig:prob_speedup}). Because \prob skips promotions randomly without distinguishing popular and unpopular objects, popular objects still have many promotions when using \prob technique. 
Moreover, the promotions of the same popular object are often performed on different CPU cores, which incurs a non-trivial number of CPU cache invalidation and coherence traffic. We conjecture that this overhead from CPU cache management limits the throughput of \prob. 
As a result, only when \probparam is very small, e.g., 0.05 or 0.01, can we observe a significant throughput increase.

When \probparam is very small, many medium-popularity objects cannot be promoted in a timely manner, which leads to a large increase in miss ratio (\cref{fig:prob_miss_ratio}). At a \probparam of 0.05, the miss ratio on 6357 traces increases by 6\% on average. Overall, \prob cannot maintain a miss ratio similar to LRU. Even a large \probparam would increase the cache miss ratio non-trivially. For example, at \probparam of 0.5, the cache miss ratio increases by 2\% on average. 

\begin{tcolorbox}[colback=black!5!white, colframe=black!75!black, left=0mm, right=0mm, top=0mm, bottom=0mm] 
\textbf{Finding.} Although \prob is very effective in reducing the number of promotions, the throughput only increases when the \probparam is very small. However, at a very small \probparam, the miss ratio increases significantly.  
\end{tcolorbox}

\begin{figure*}[ht]
    \centering
    \begin{subfigure}[b]{0.28\linewidth}
        \centering
        \includegraphics[width=\linewidth]{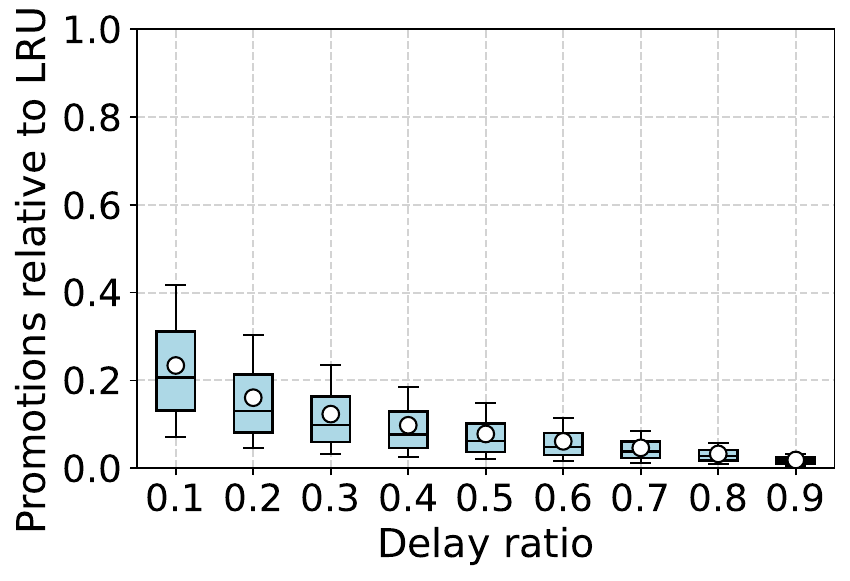}
        \caption{Relative number of promotions}
        \label{fig:delay_promotions}
    \end{subfigure}
   \hspace{0.8em}
    \begin{subfigure}[b]{0.28\linewidth}
        \centering
        \includegraphics[width=\linewidth]{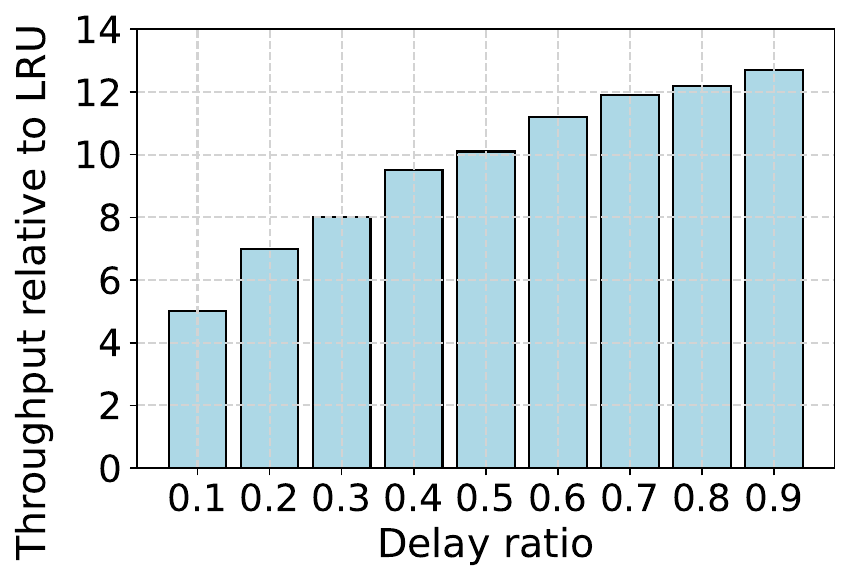}
        \caption{Throughput Imprv (16 threads)}
        \label{fig:delay_speedup}
    \end{subfigure}
   \hspace{0.8em}
    \begin{subfigure}[b]{0.29\linewidth}
        \centering
        \includegraphics[width=\linewidth]{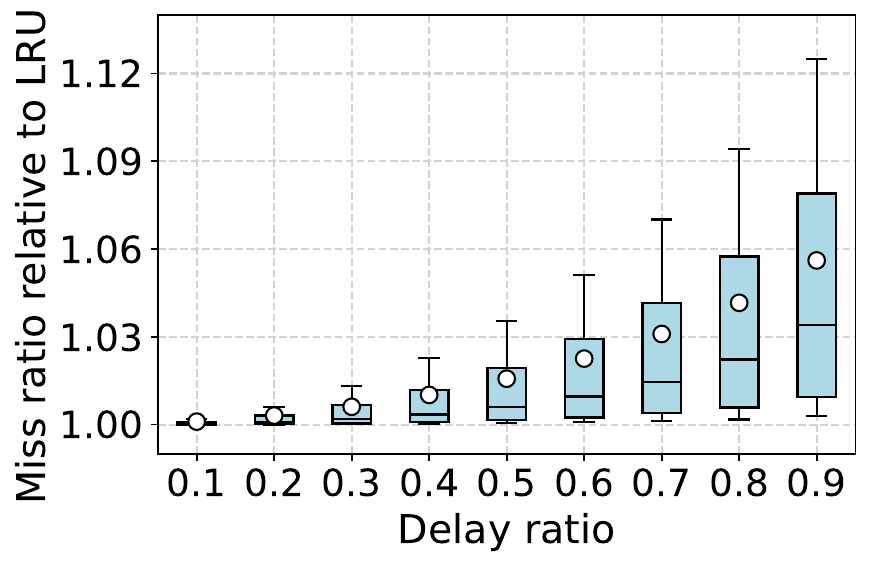}
        \caption{Relative miss ratio}
        \label{fig:delay_miss_ratio}
    \end{subfigure}
    \caption{Delay-LRU reduces cache promotions by skipping recently-promoted objects. \normalfont{When the \delayparam is 20\% of the cache size, \delay increases LRU's miss ratio by no more than 0.1\% while reducing the number of promotions by over 82\% and increasing throughput by 7$\times$. }}
    \label{fig:delay_miss_speedup_promotion}
\end{figure*}

\subsection{\batch} \label{sec:batch}
\begin{lstlisting}[caption={promotion during cache hit in \batch}, style=custom, language=Python, label=batchListing]
batch_time = batch_ratio * cache_size
batch.insert(obj)
if time - last_prom_time >= batch_time:
    for item in batch:
        if item in cache:
            promote(item)
    batch.clear()
\end{lstlisting}

Because the locking operation at each promotion is computationally expensive~\cite{ding_bp-wrapper_2009} and leads to contention, 
\batch performs promotions in batches to reduce the overhead and contention. Instead of immediately moving the requested object to the head of the linked list, \batch tracks the requested object IDs in a buffer and periodically promotes them. For 8-byte object id, a batch of 128 consumes only 1~KB, resulting in negligible memory overhead.

The efficiency and performance of \batch depend on one parameter: \batchparam, which decides how often \batch performs a batched promotion. For the sake of consistency in this paper, we measure the time between promotions using insertion time, which increments by one upon each insertion into the cache. Therefore, a \batchparam of 0.1 means that there are $0.1\times$\textit{cache size} insertions between each batch promotion. The details can be found in \Cref{batchListing}.

Batching can effectively reduce the number of promotions because multiple promotions of a popular object are combined into a single promotion; however, the reduction depends heavily on the workload. \Cref{fig:batch_promotions} shows that the variance in promotion reduction is highly variable. Recall that each box shows the 25th and 75th percentiles of the 6357 traces. The long box indicates that some traces enjoy a huge promotion reduction, while others do not observe much benefit. For example, at a small \batchparam of 0.1, the number of promotions on the mean and median trace can drop to 60\% of that in LRU, while the worst trace does not benefit from batched promotion at all. We find that \textit{the effectiveness of \batch depends on the skewness of the workload: a more skewed workload tends to exhibit larger benefits} because more requests are for a few popular objects. Therefore, more requests in a batch are coalesced into a single request. 

Although \batch shows a large variance in promotion reduction, it is effective in improving throughput when the number of promotions can be reduced. \Cref{fig:batch_speedup} shows that even at a small \batchparam of 0.1, the throughput increases by more than 4$\times$ at 16 threads. Further increasing the \batchparam leads to continued throughput increase. Compared to \prob, \batch provides better scalability because \batch reduces more promotions from popular objects while \prob reduces promotion for popular and unpopular objects with equal probability.  

Besides being an effective scalability improvement technique, \batch is also more effective in maintaining a low miss ratio. \Cref{fig:batch_miss_ratio} shows that when the \batchparam is small, e.g., no more than 0.5, the miss ratio increases for a median trace by less than 1\%. The reason is that \batch prioritizes reducing the promotion of popular objects, which are less likely to cause an increase in the miss ratio. When \batchparam is very large, some objects in the batch may have been evicted before they are promoted, leading to an increased miss ratio. 

\begin{tcolorbox}[colback=black!5!white, colframe=black!75!black, left=0mm, right=0mm, top=0mm, bottom=0mm] 
\noindent \textbf{Finding.} \batch is an effective \lp technique as it can significantly reduce the number of promotions, leading to better scalability. Meanwhile, \batch can maintain a miss ratio similar to LRU. However, the effectiveness of \batch is highly variable and heavily depends on the workload. 
\end{tcolorbox}

\subsection{\delay} \label{sec:delay}
\begin{lstlisting}[caption={promotion during cache hit in \delay}, style=custom, language=Python, label=delayListing]
delay_time = delay_ratio * cache_size
if time - obj.last_prom_time > delay_time:
    promote(obj)
    obj.last_prom_time = current_time
\end{lstlisting}

Upon a cache hit, \delay checks and promotes only if sufficient time has elapsed since the object's last promotion.
Although this approach requires storing a timestamp for each object, a 4-byte timestamp adds negligible 0.1\% overhead to typical 4~KB objects.
Similar to \Cref{sec:batch}, the delay time is measured using the number of insertions. \delayparam is the ratio between delay time and cache size, ranging from 0 to 1. A \delayparam of 1 would cause every object to be evicted before promotion. 
Upon a cache hit, we compare the last promotion time of the object and the current time to decide whether the object should be promoted. 

Compared to \prob and \batch, \delay is very effective at reducing the number of promotions. 
\Cref{fig:delay_promotions} shows that when \delayparam is 10\% of cache size, the number of promotions is reduced to 24\% of that in LRU on average without visibly increasing miss ratio. This indicates that around 76\% of promotions occur soon after the previous request to the same object, and \delay can avoid these unnecessary promotions. 

\begin{figure*}[ht]
    \centering
    \begin{subfigure}[b]{0.28\linewidth}
        \centering
        \includegraphics[width=\linewidth]{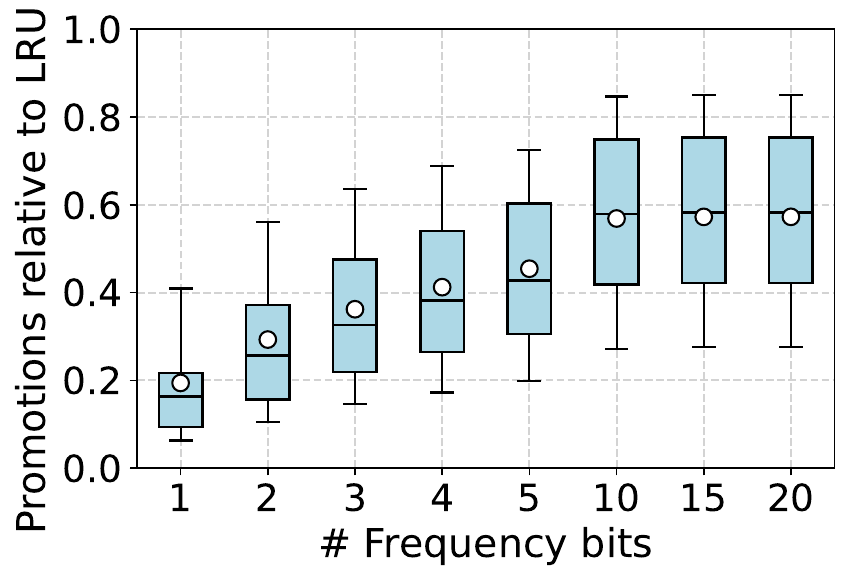}
        \caption{Relative number of promotions}
        \label{fig:clock_promotions}
    \end{subfigure}
   \hspace{0.8em}
    \begin{subfigure}[b]{0.28\linewidth}
        \centering
        \includegraphics[width=\linewidth]{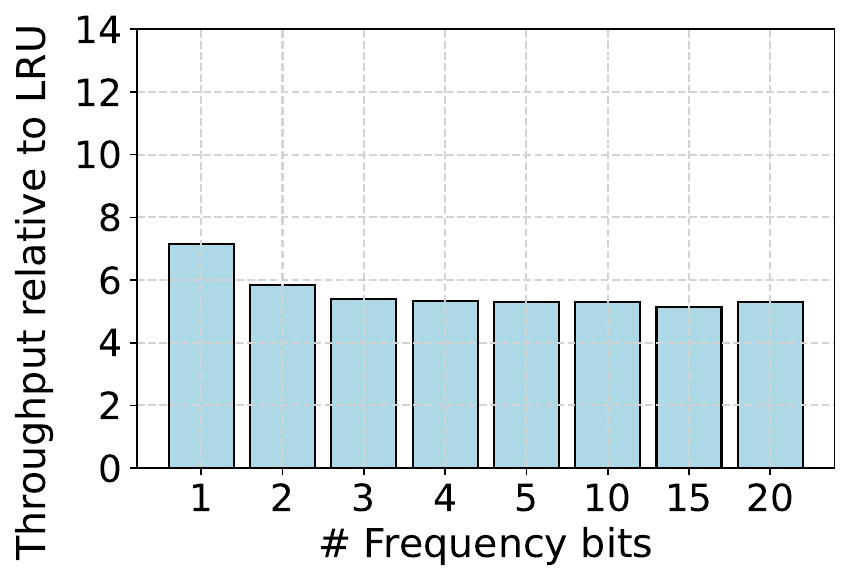}
        \caption{Throughput Imprv (16 threads)}
        \label{fig:clock_speedup}
    \end{subfigure}
   \hspace{0.8em}
    \begin{subfigure}[b]{0.29\linewidth}
        \centering
        \includegraphics[width=\linewidth]{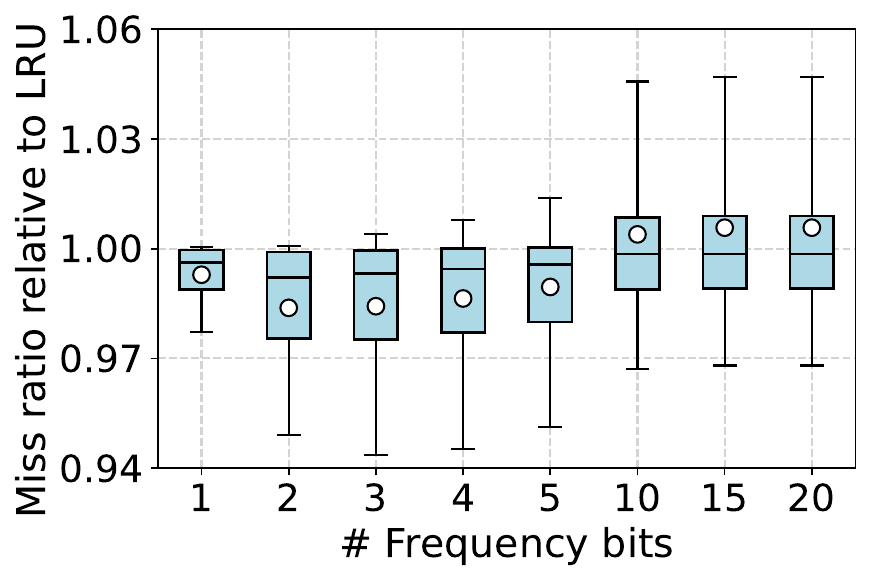}
        \caption{Relative miss ratio}
        \label{fig:clock_miss_ratio}
    \end{subfigure}
    \caption{\clock reduces both promotions and cache misses. \normalfont{The reduced promotions translate to a higher throughput. However, the scalability upper bound is lower than other \lp techniques such as \delay.}}
    \label{fig:clock_miss_speedup_promotion}
\end{figure*}

The significant reduction in the number of promotions allows \delay to achieve notable scalability gains. At the 0.1 \delayparam, \delay exhibits a 5$\times$ higher throughput than LRU at 16 threads (\Cref{fig:delay_speedup}). As the \delayparam increases, the speedup further increases and can achieve more than 12$\times$ throughput speedup at a 0.9 \delayparam. However, such a scalability improvement comes with a cost on efficiency---the long \delayparam increases the miss ratio by 6\% on average on 6357 traces as shown in \Cref{fig:delay_miss_ratio}. 
Although a large \delayparam leads to a high miss ratio, when the \delayparam is small, the increase in the miss ratio is negligible. For example, at a \delayparam of 0.1, adding delayed promotion to LRU does not increase the miss ratio for 39.5\% of the traces. 
This suggests that although promotions are helpful for LRU in achieving a low miss ratio, frequently promoting popular objects does not bring much benefit, and we can skip these promotions without hurting the miss ratio. 

Both \batch and \delay prioritize reducing the promotions of popular objects; however, \delay is more effective and offers consistent benefits compared to \batch. For example, at a \delayparam of 0.1, \delay achieves almost the same miss ratio as LRU with a 5$\times$ higher throughput. To achieve a similar throughput, \batch needs to use a \batchparam of 0.2, which has a miss ratio 0.2\% higher than LRU on average. Moreover, if we prefer a higher throughput, \delay can provide more than 12$\times$ LRU's throughput, while \batch can only achieve at most 8.2$\times$ LRU's throughput~\footnote{Note that because \delayparam and \batchparam are different, we cannot compare the throughput and miss ratio of the same \delayparam and \batchparam. We should only compare them under the same miss ratio or throughput.}. The reason behind \batch's lower throughput is that \batch needs to perform more work in the critical section. 


\begin{tcolorbox}[colback=black!5!white, colframe=black!75!black, left=0mm, right=0mm, top=0mm, bottom=0mm] 
\noindent \textbf{Finding.} \delay can effectively improve LRU's scalability with a minimal impact on efficiency. Its effectiveness remains consistent across diverse workloads. 
\end{tcolorbox}

\subsection{\clock} \label{sec:clock}
\begin{lstlisting}[caption={\clock performs promotion (reinsertion) during cache eviction.}, style=custom, language=Python, label=clockListing]
obj_to_evict = queue.front()
while obj_to_evict.freq > 0:
    obj_to_evict.freq -=1
    next_obj = obj_to_evict.next
    promote(obj_to_evict)
    obj_to_evict = next_obj
evict(obj_to_evict)
\end{lstlisting}

\clock, as the name suggests, uses a FIFO queue to order objects, and there is no change to the ordering during cache hits. During cache evictions, popular objects at the tail are reinserted into the cache. Unlike other \lp techniques that reduce the number of promotions at each cache hit, \clock delays the ``promotion'' to eviction time. Therefore, we count the number of reinsertions as promotions. 
 
Each object in \clock is associated with a counter that tracks the object's access frequency. The counter is capped based on the number of bits used. For example, a \clockparam of 1 caps the maximum frequency at 1, effectively acting as a boolean variable. A \clockparam of 3 caps the frequency at 7: objects accessed more than 7 times in the cache are treated as 7. Upon a cache hit, the counter increments by 1. During evictions, objects with a counter larger than 0 are reinserted with the counter decremented by 1, and objects with a frequency counter of 0 are evicted. 

\Cref{fig:clock_promotions} shows that the number of promotions (reinsertions) can significantly reduce in \clock, especially when the number of \clockparam is small. When the number of \clockparam increases, the number of promotions increases until it reaches a plateau. This occurs because, after a certain threshold, increasing the frequency cap further does not affect the reinserted objects, and it is always the same set of objects that are being reinserted. 

Similar to \delay and \batch, most of the promotion reduction comes from popular objects. As a result, \Cref{fig:clock_speedup} shows that \clock also enjoys a significant boost in throughput. However, the highest throughput \clock can achieve is lower than \delay. Because \delay can reduce the number of promotions by more than 95\%, it can increase throughput by more than 12$\times$. In contrast, \clock can only reduce the number of promotions by 80\% at \clockparam 1. 

Although \clock is slightly worse than \delay in terms of scalability, it is more efficient. \emph{It is the only \lp technique that reduces LRU's miss ratio}. \Cref{fig:clock_miss_ratio} shows that with a 2-bit frequency counter, \clock reduces LRU's miss ratio by almost 2\% on average. Increasing the number of bits in the frequency counter initially reduces the miss ratio but eventually causes it to rise. This occurs because a large frequency cap causes some short-lived popular objects to become stuck in the cache. 
In such workloads, hot objects frequently change, but their high frequencies prevent them from being evicted quickly. 

\begin{tcolorbox}[colback=black!5!white, colframe=black!75!black, left=0mm, right=0mm, top=0mm, bottom=0mm] 
\textbf{Finding.} 
\clock is most effective when we use 1-bit or 2-bit counters. It not only improves scalability, but also improves cache efficiency. 
\end{tcolorbox}

\subsection{\random} \label{sec:random}


\begin{figure}[t]
    \centering
    \begin{subfigure}[b]{0.48\linewidth}
        \centering        \includegraphics[width=\linewidth]{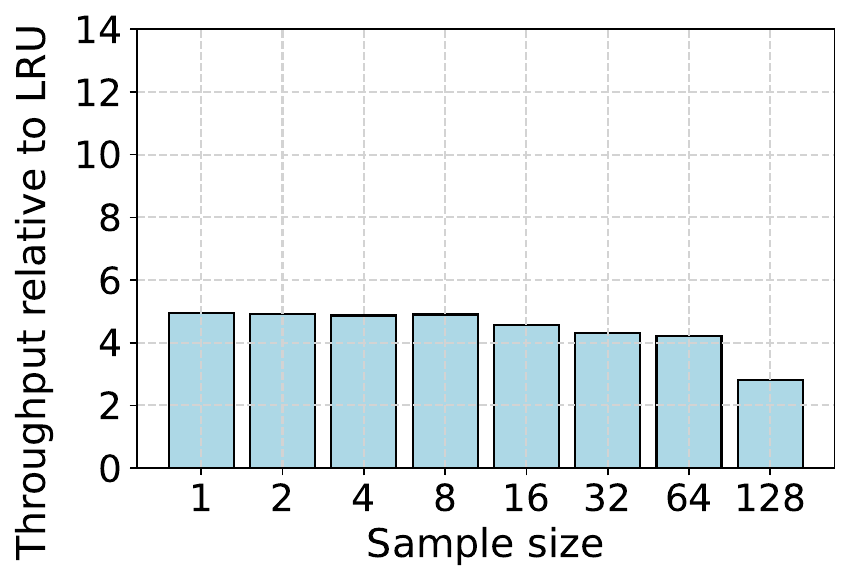}
        \caption{Throughput improvement}
        \label{fig:random_speedup}
    \end{subfigure}
    \begin{subfigure}[b]{0.49\linewidth}
        \centering
        \includegraphics[width=\linewidth]{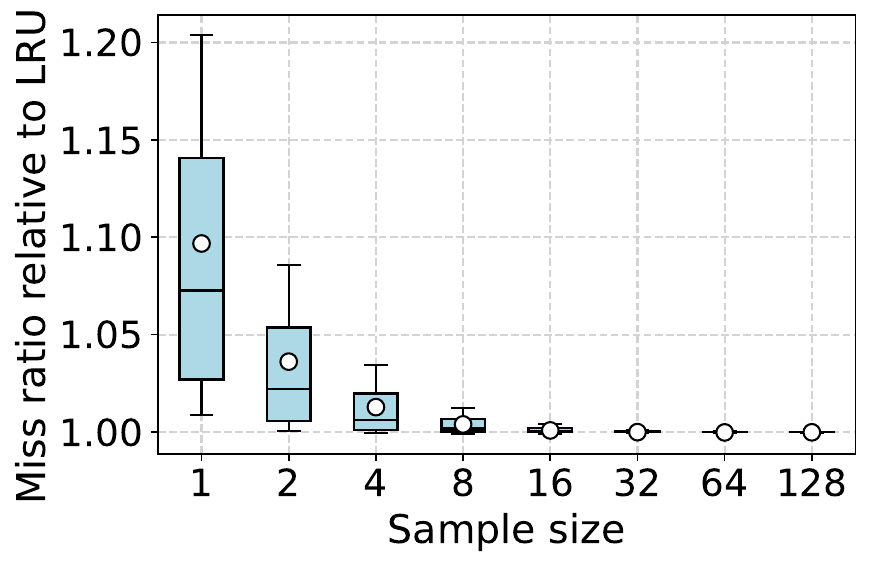}
        \caption{Relative miss ratio}
        \label{fig:random_miss_ratio}
    \end{subfigure}
    \caption{\random samples objects and evicts the least-recently-used one during eviction. \normalfont{Removing the global linked list allows it to be more scalable than LRU. However, it underperforms LRU in efficiency. }}
    \label{fig:random_miss_speedup}
\end{figure}

\begin{figure*}[ht]
    \centering
    \begin{subfigure}[b]{0.24\linewidth}
        \centering
        \includegraphics[width=\linewidth]{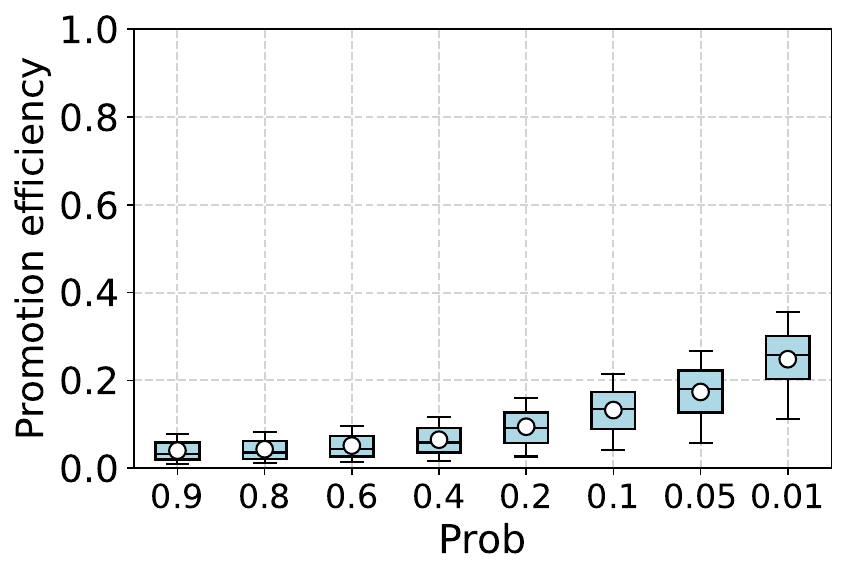}
        \caption{\prob}
        \label{fig:prob_promotion_efficiency}
    \end{subfigure}
    \begin{subfigure}[b]{0.24\linewidth}
        \centering
        \includegraphics[width=\linewidth]{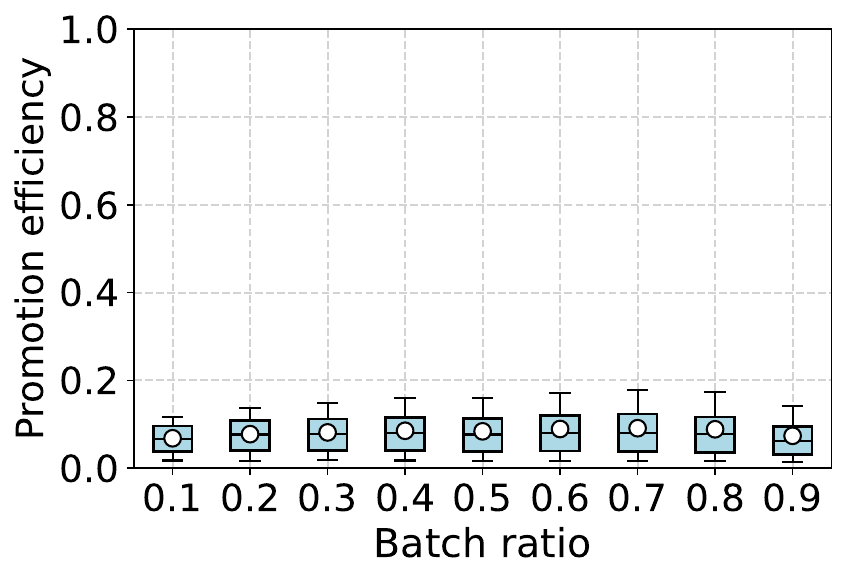}
         \caption{\batch}
        \label{fig:batch_promotion_efficiency}
    \end{subfigure}
    \begin{subfigure}[b]{0.24\linewidth}
        \centering
        \includegraphics[width=\linewidth]{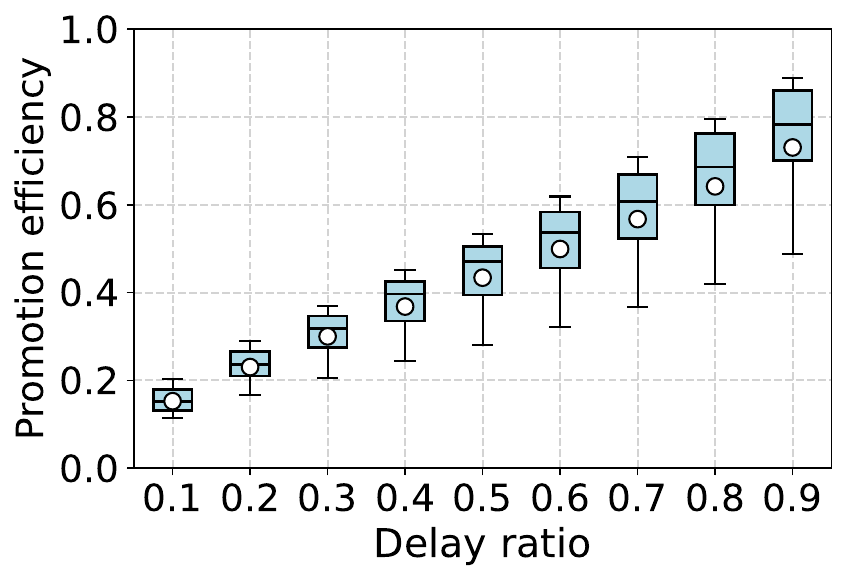}
         \caption{\delay}
        \label{fig:delay_promotion_efficiency}
    \end{subfigure}
    \begin{subfigure}[b]{0.24\linewidth}
        \centering
        \includegraphics[width=\linewidth]{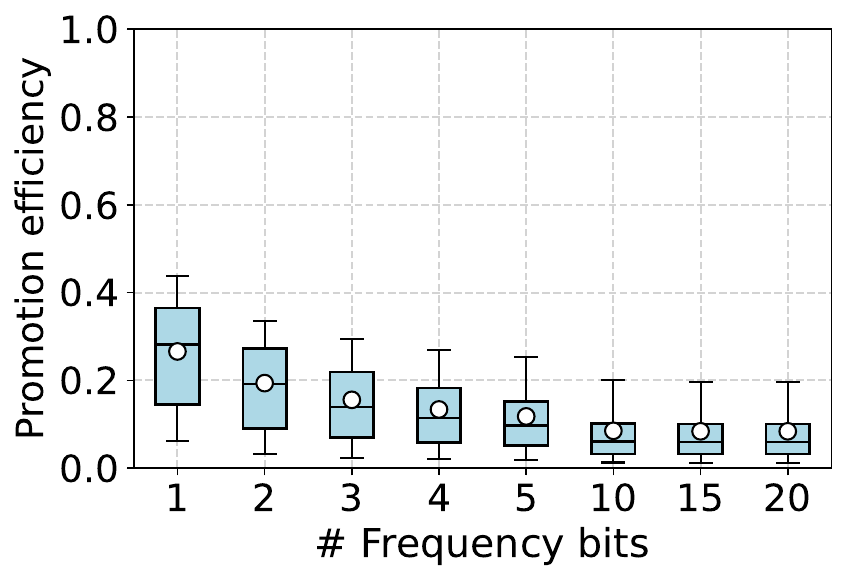}
         \caption{\clock}
        \label{fig:clock_promotion_efficiency}
    \end{subfigure}
    \vspace{-1.2em}
    \caption{Promotion efficiency calculated as the number of misses reduced from FIFO over the number of promotions. Higher is better. }
    \label{fig:promotion_effciency}
\end{figure*}

\random samples some objects from the cache during eviction and evicts the least-recently used sample. It is not a \lp promotion technique; however, it can also improve scalability. Therefore, we add it to the discussion. 

Because \random does not maintain a global linked list and there is no locking during cache hits, it is more scalable than LRU. \Cref{fig:random_speedup} shows that \random can achieve a throughput of 5$\times$ higher than LRU. However, the improvement is smaller compared to \delay due to the overhead of random sampling and locking during eviction. This overhead increases with the number of samples. We can see that the throughput reduces to 3$\times$ that of LRU when \random samples 128 objects at each eviction. 

While \random improves scalability, \Cref{fig:random_miss_ratio} shows that it is lackluster in efficiency. 
Using just one sample, it becomes the random eviction algorithm without using recency information. We observe that it increases the miss ratio by 10\% on average on the 6357 traces. Increasing the number of samples reduces the miss ratio. With 16 samples per eviction, we find that \random achieves a similar miss ratio to LRU. 
Conventional wisdom suggests that \random can achieve a miss ratio lower than LRU for workloads that exhibit loop access patterns because such a pattern causes thrashing for LRU when the loop size is larger than the cache size. Surprisingly, this observation does not appear in our evaluations. Although over 70\% of the 6357 traces are block cache workloads, which tend to have loop and scan access patterns, we find LRU to be better than \random on less than 12\% of the traces under different sample sizes. 

\begin{tcolorbox}[colback=black!5!white, colframe=black!75!black, left=0mm, right=0mm, top=0mm, bottom=0mm] 
\textbf{Finding.} 
\random can improve scalability without compromising the miss ratio but requires careful choice of parameters. Using 16 samples per eviction provides a good balance. Increasing the sample size reduces throughput, while reducing the sample size increases the miss ratio. 
\end{tcolorbox}

\subsection{Promotion efficiency}

We have shown that different \lp techniques perform differently. \prob can significantly reduce the number of promotions with the cost of increasing miss ratio; \batch heavily depends on the workload---very effective in reducing the number of hits for some workloads without increasing miss ratio; \delay performs the best in reducing the number of promotions while maintaining miss ratio; \clock, compared to other \lp techniques, not only reduces the number of promotions but also reduces miss ratio. 

To further characterize the effectiveness of promotion, we propose a new metric called ``promotion efficiency'', which calculates the misses that each promotion reduces from FIFO on average. If a technique or algorithm can significantly reduce FIFO's miss ratio with very few promotions, then it has a high promotion efficiency. Since FIFO has no promotion, it is not included in the figure. LRU promotes upon each cache hit, most of which are unnecessary, resulting in an average promotion efficiency of 0.037. 

\prob has a promotion efficiency similar to LRU when \probparam is high (\Cref{fig:prob_promotion_efficiency}). However, when \prob drops to 0.01, we observe an increase in promotion efficiency. This happens because when the probability of promotion is very low, only very popular objects may be promoted. 
Unlike FIFO, which does not keep popular objects in the cache, these occasional promotions of very popular objects help keep them in the cache, driving up promotion efficiency. 

\Cref{fig:batch_promotion_efficiency} shows the promotion efficiency of \batch. We find that across different \batchparams, promotion efficiency is consistently low and not sensitive to \batchparam. This occurs because \batch's effectiveness depends on the workload and cannot significantly reduce the number of promotions for many traces. 

Compared to \prob and \batch, \Cref{fig:delay_promotion_efficiency} shows that \delay can achieve remarkably high promotion efficiency. At a \delayparam of 0.4, each promotion in \delay can reduce 0.4 miss from FIFO. Further increasing the \delayparam to 0.9, \delay can achieve the highest mean promotion efficiency among all \lp techniques at 0.72. Because \delay reduces the promotion of popular objects, which is very effective at identifying and reducing the most useless promotions. 

\Cref{fig:clock_promotion_efficiency} shows that \clock has the highest promotion efficiency when using a 1-bit counter. Using higher frequency caps only reduces the promotion efficiency. This is because using more bits to record frequency often leads to unnecessary promotions. 

Among all the \lp techniques, we find that \delay has the most efficient promotions. \clock closely follows \delay when using a 1-bit counter. \batch and \prob have the lowest promotion efficiency, and many promotions do not reduce cache misses.

\subsection{\lp on Advanced Eviction Algorithms}
\begin{figure*}[ht]
    \centering
    \begin{subfigure}[b]{0.24\linewidth}
        \centering
        \includegraphics[width=\linewidth]{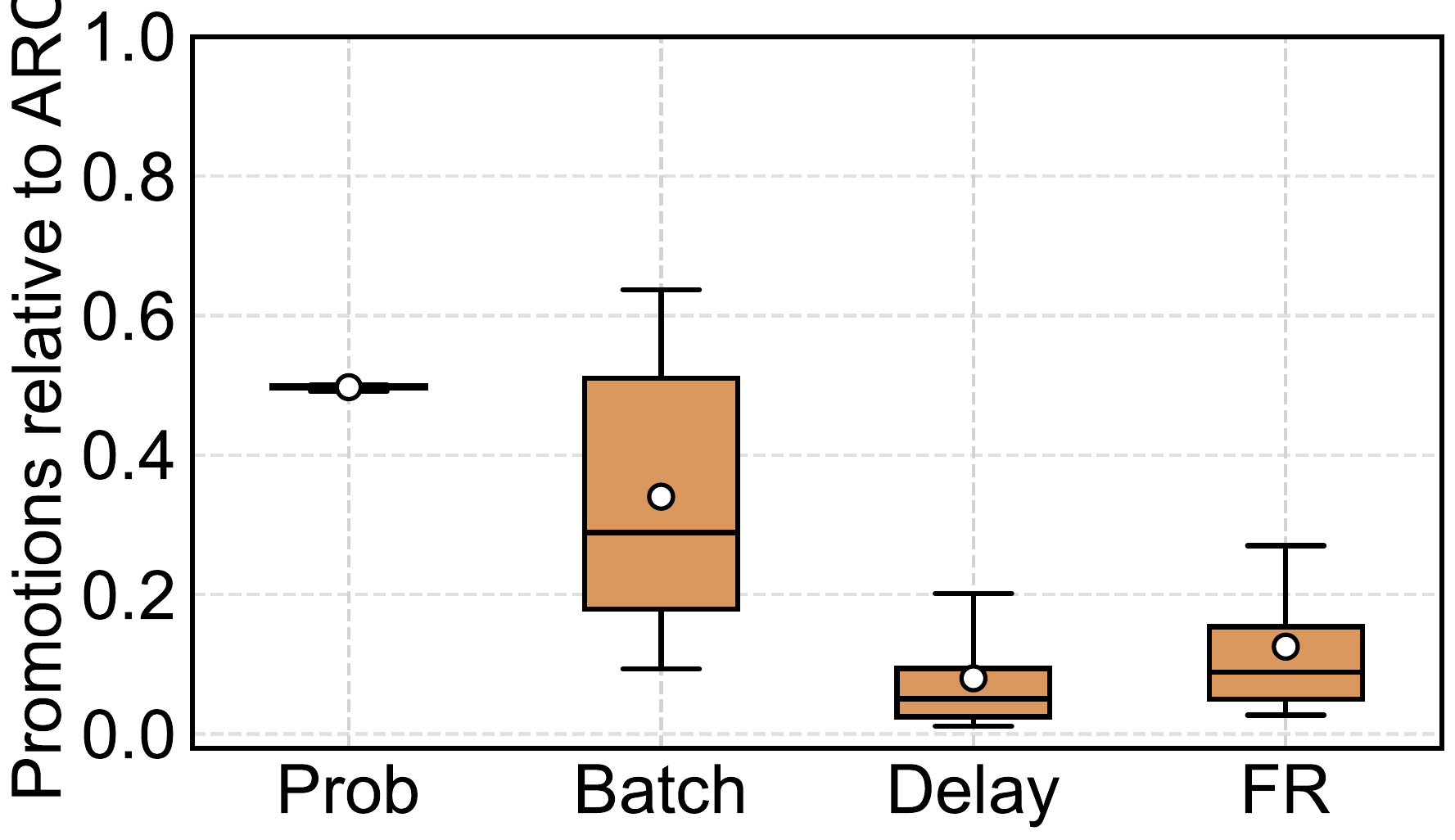}
        \caption{ARC}
        \label{fig:advanced_arc_promotion}
    \end{subfigure}
    \begin{subfigure}[b]{0.24\linewidth}
        \centering
        \includegraphics[width=\linewidth]{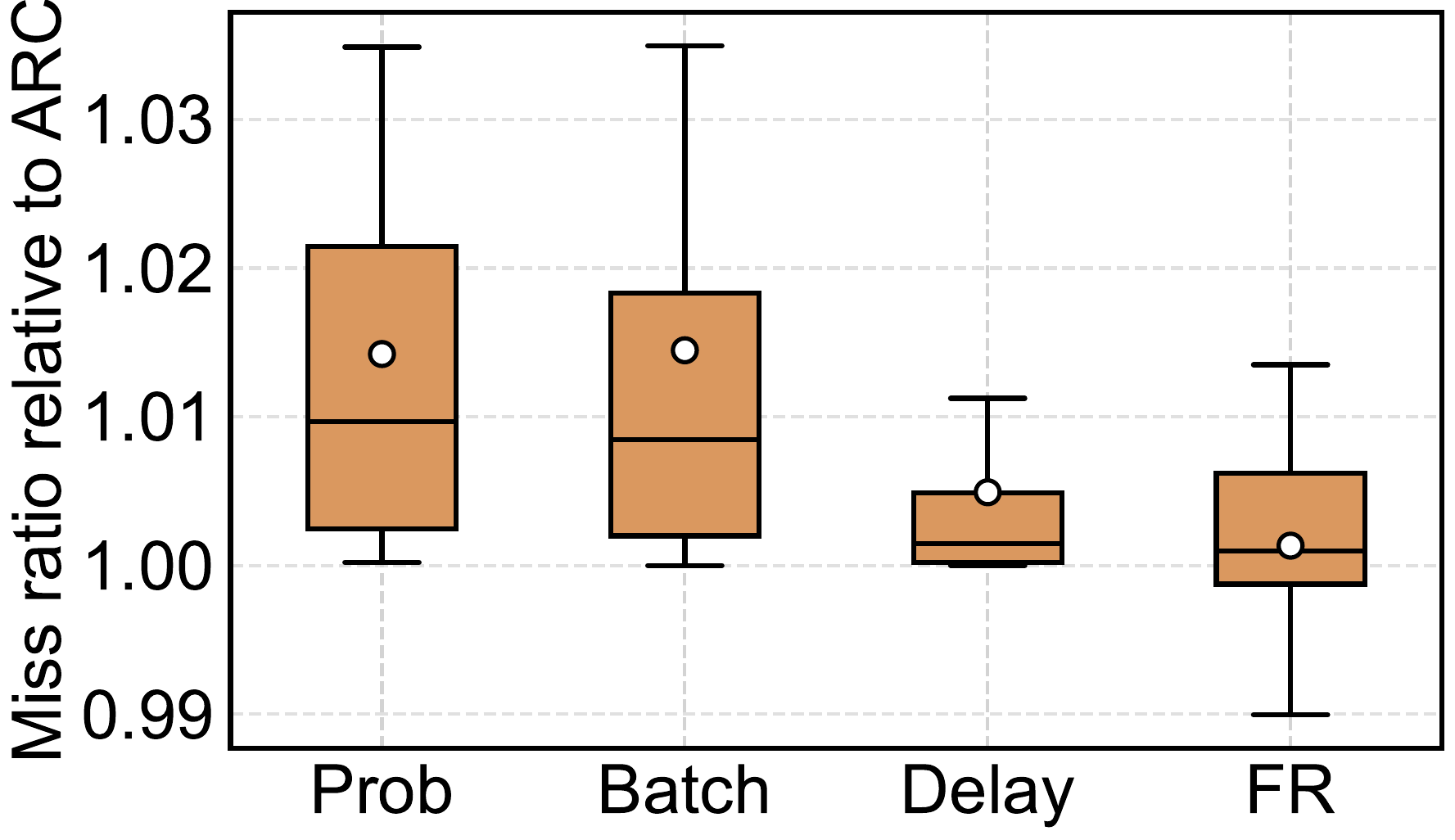}
         \caption{ARC}
        \label{fig:advanced_arc_miss}
    \end{subfigure}
    \begin{subfigure}[b]{0.24\linewidth}
        \centering
        \includegraphics[width=\linewidth]{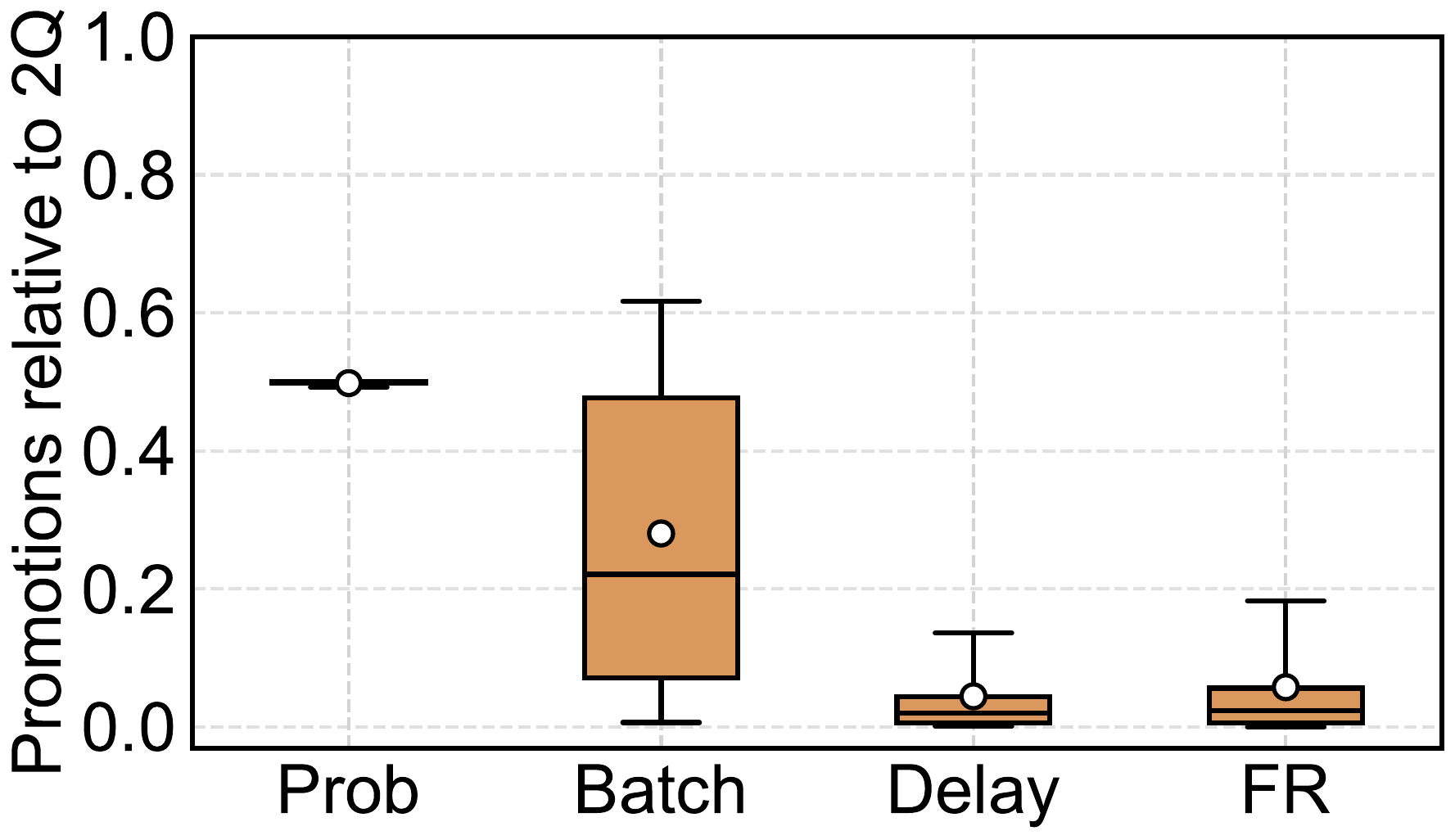}
         \caption{2Q}
        \label{fig:advanced_twoq_promotion}
    \end{subfigure}
    \begin{subfigure}[b]{0.24\linewidth}
        \centering
        \includegraphics[width=\linewidth]{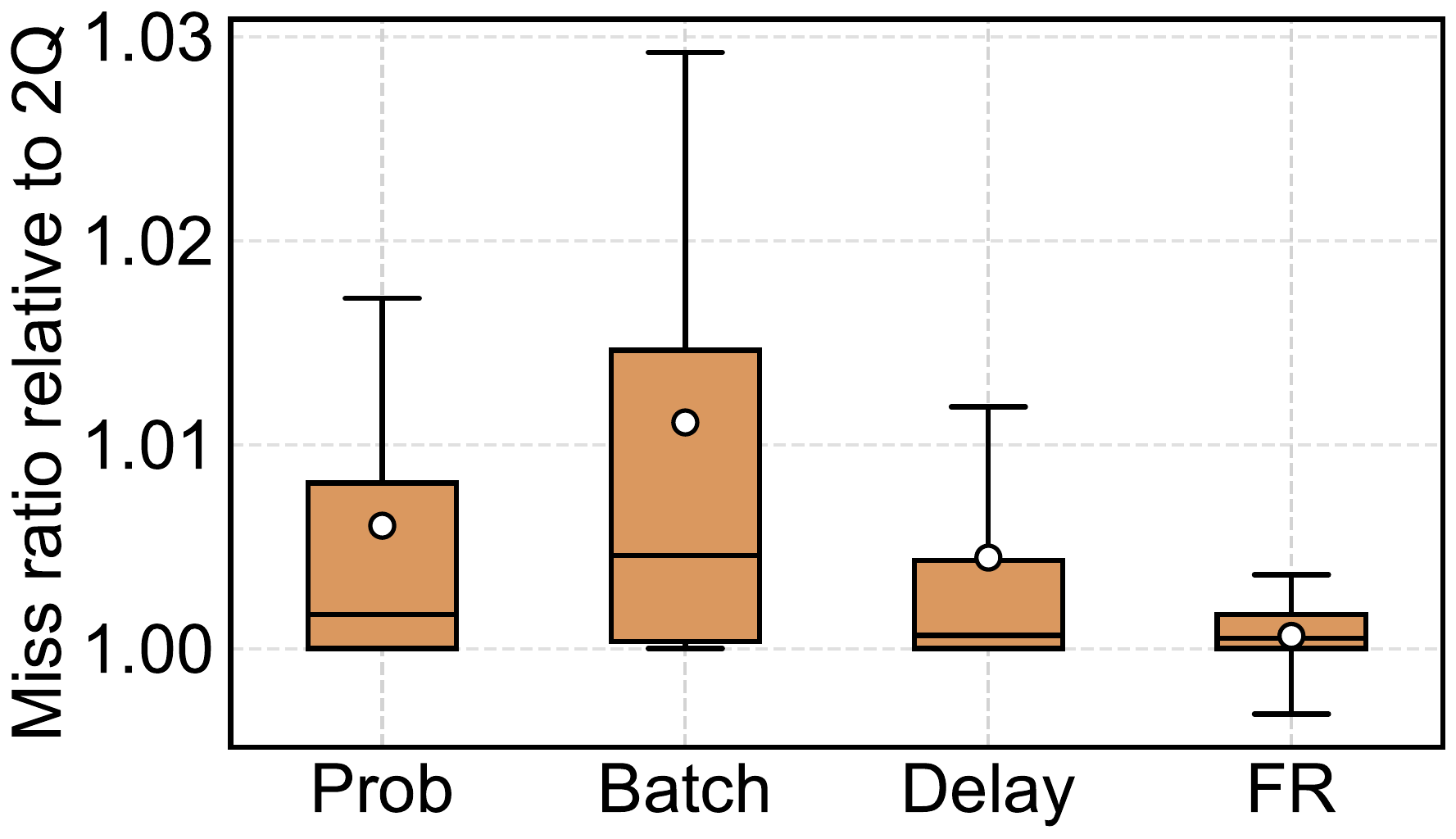}
         \caption{2Q}
        \label{fig:advanced_twoq_miss}
    \end{subfigure}
    \vspace{-1.2em}
    \caption{Using \lp in advanced LRU-based eviction algorithms. \normalfont{Similar to LRU, delay and reinsertion are more effective than batch and probabilistic promotion. }}
    \label{fig:advanced}
\end{figure*}

Most advanced cache eviction algorithms designed in the past two decades are built on top of one or more LRU queues using different promotion metrics. For example, ARC\cite{megiddo_arc_2003}, SLRU\cite{huang_analysis_2013}, 2Q\cite{johnson_2q_1994}, MQ\cite{zhou_multi-queue_2001}, and multi-generational LRU\cite{osawa1997generational} leverage multiple LRU queues to differentiate between frequently and infrequently accessed items. Others, such as LIRS\cite{jiang_lirs_2002} and LIRS2\cite{zhong_lirs2_2021}, retain an LRU queue but apply different metrics, e.g., stack distance, to determine how to promote an object. 
Because these advanced algorithms are built on top of LRU, they also suffer from limited scalability. The \lp techniques studied in this paper can also be used to improve their scalability. 
We illustrate this with 2Q~\cite{johnson_2q_1994} and ARC~\cite{megiddo_arc_2003}.

Both 2Q and ARC comprise LRU queues that transition objects from ``recent'' to ``frequent'' states to improve miss ratio (A$_1$ to A$_m$ in 2Q; T$_1$ to T$_2$ in ARC). 
We refrain from interfering with the policy decision to enter the frequent queue. We add \lp to the LRU queue for frequent objects in each algorithm, i.e., within A$_m$ (2Q) and T$_2$ (ARC), to regulate in-queue promotions. 

We have examined the impact of parameters on each \lp technique in this section. We will use 0.5 for the rest of the sections so that we only have one box for each technique.  
\Cref{fig:advanced} shows that ARC and 2Q exhibit patterns similar to those observed on LRU: delay and reinsertion are more effective than batch and probabilistic promotion. Delay can significantly reduce promotions without significantly increasing the miss ratio, while reinsertion can reduce both the miss ratio and the number of promotions. Moreover, the effectiveness of batching promotions is workload-dependent---some workloads benefit more, while others benefit less.  
These results indicate that LP’s benefits are not artifacts of a single policy, but rather arise from suppressing low-value promotions that are common in LRU-style designs.

\section{Can lazy promotions be lazier with future information?} \label{sec:lp_lessons}

\begin{figure*}[!ht]
    \centering
    
    \begin{subfigure}[b]{0.28\linewidth}
        \centering
        \includegraphics[width=\linewidth]{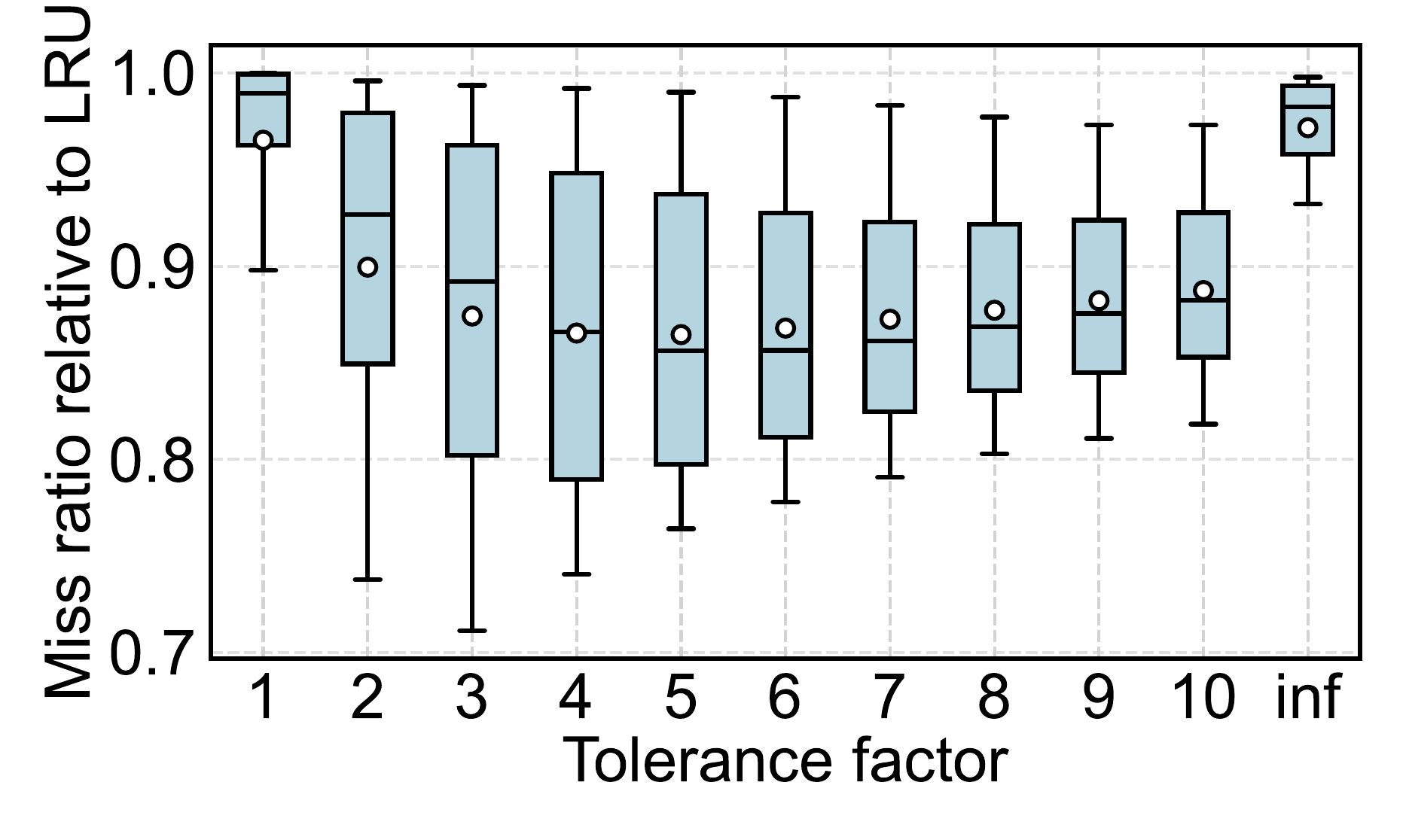}
        \caption{Miss ratio of \beladyrandomLRU }
        \label{fig:rbl_0.01}
    \end{subfigure}
    \hspace{0.8em}
     \begin{subfigure}[b]{0.28\linewidth}
        \centering
        \includegraphics[width=\linewidth]{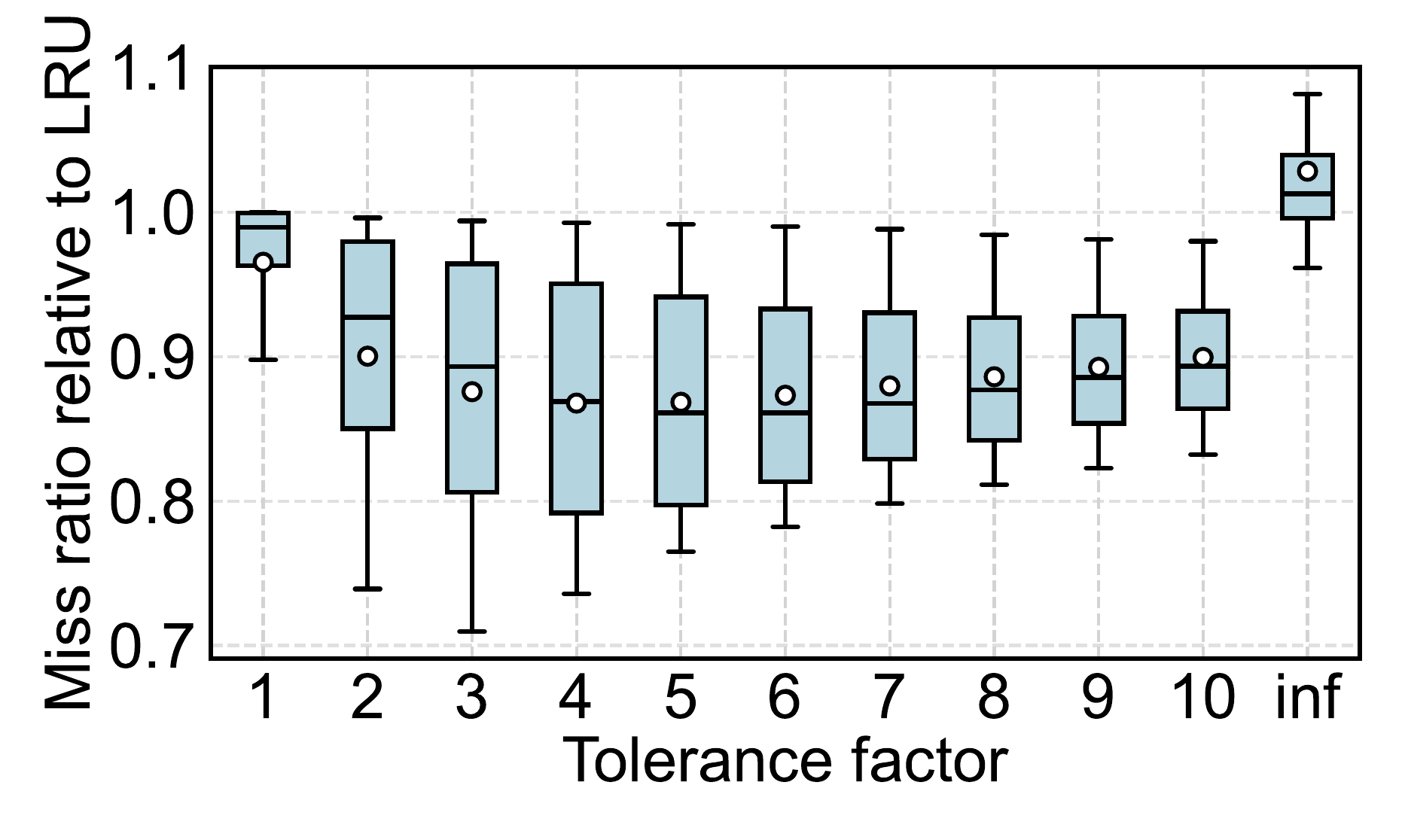}
        \caption{Miss ratio of \beladyrandom}
        \label{fig:rb_0.01}
    \end{subfigure}
    \hspace{0.8em}
    \begin{subfigure}[b]{0.27\linewidth}
        \centering
        \includegraphics[width=\linewidth]{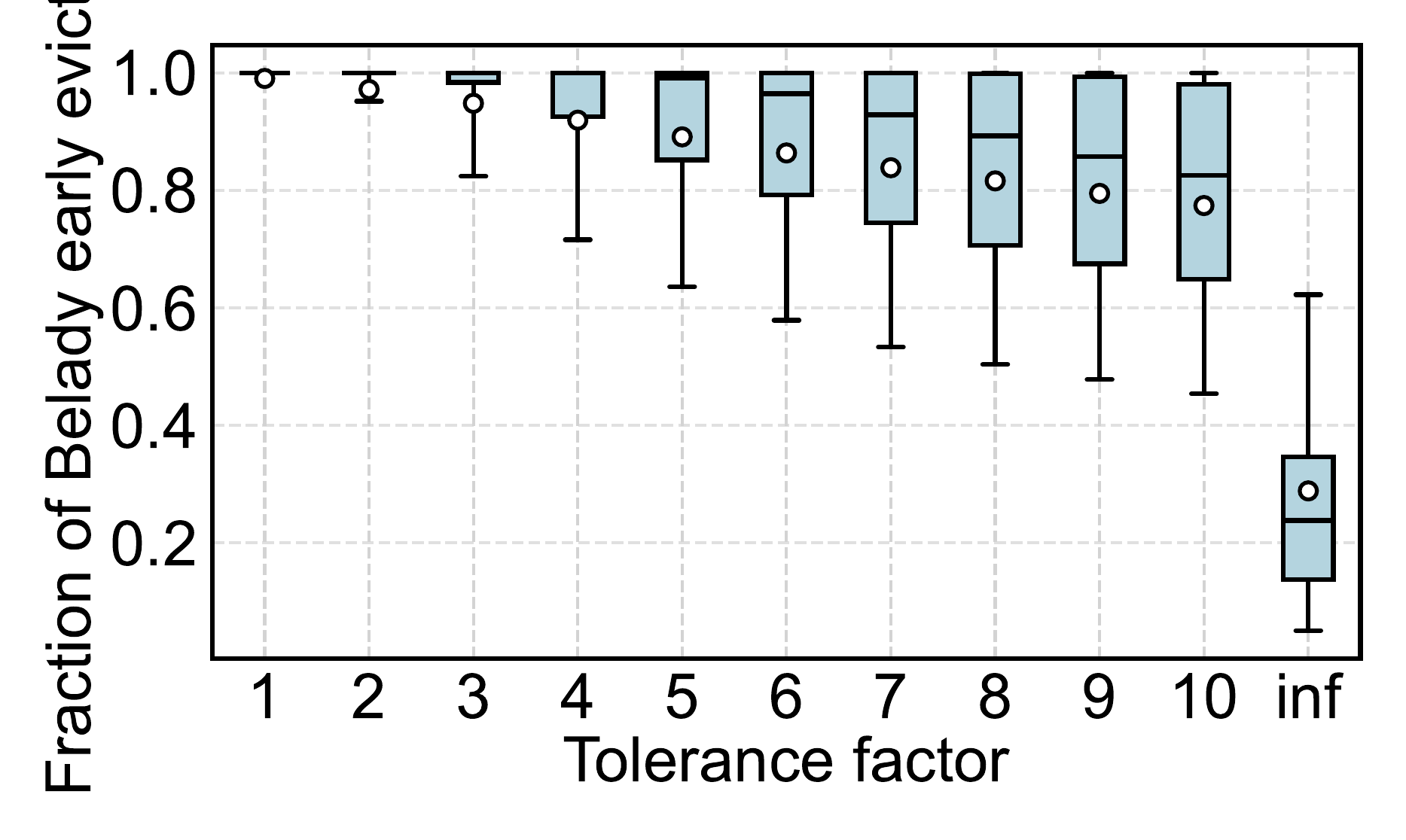}
        \caption{Fraction of Belady early evictions}
        \label{fig:distribution_0.01}
    \end{subfigure}
    \vspace{-1.2em}
    \caption{Using future access time to early evict objects (denoted as Belady early eviction) allows a cache to avoid ranking (promoting) objects. \normalfont{a) and b) when Belady early eviction does not evict enough objects, \random and Random are used for eviction, respectively. We find that \random and Random do not show much difference in miss ratio, suggesting that ranking objects in the cache for eviction is unnecessary. c) The reason why \beladyrandomLRU and \beladyrandom show similar miss ratios is that Belady early eviction accounts for most of the cache evictions. }}
    \label{fig:rb-alg}
\end{figure*}

\begin{figure*}[ht]
\centering
    \begin{subfigure}[b]{0.28\linewidth}
        \includegraphics[width=\linewidth]{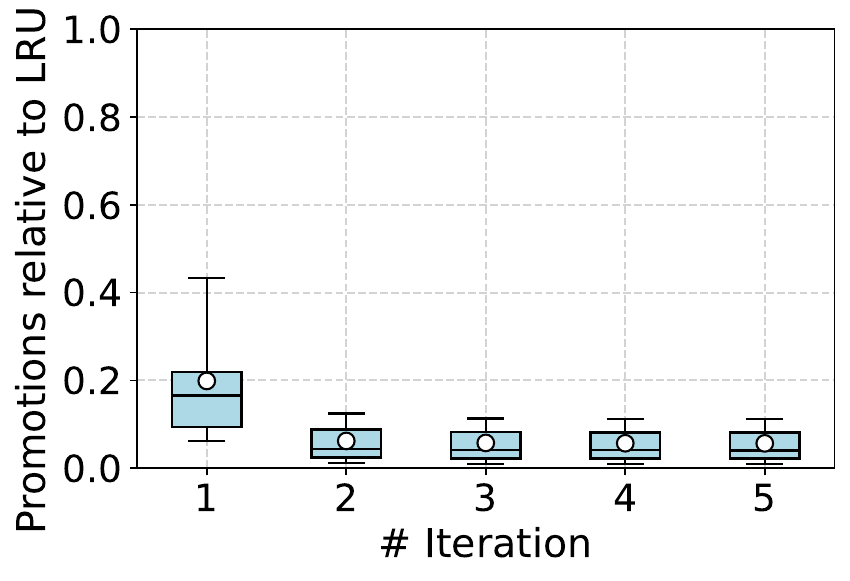}
        \caption{Relative number of promotions}
        \label{fig:bc_promotion}
    \end{subfigure}
    \hspace{0.8em}
    \begin{subfigure}[b]{0.28\linewidth}
        \includegraphics[width=\linewidth]{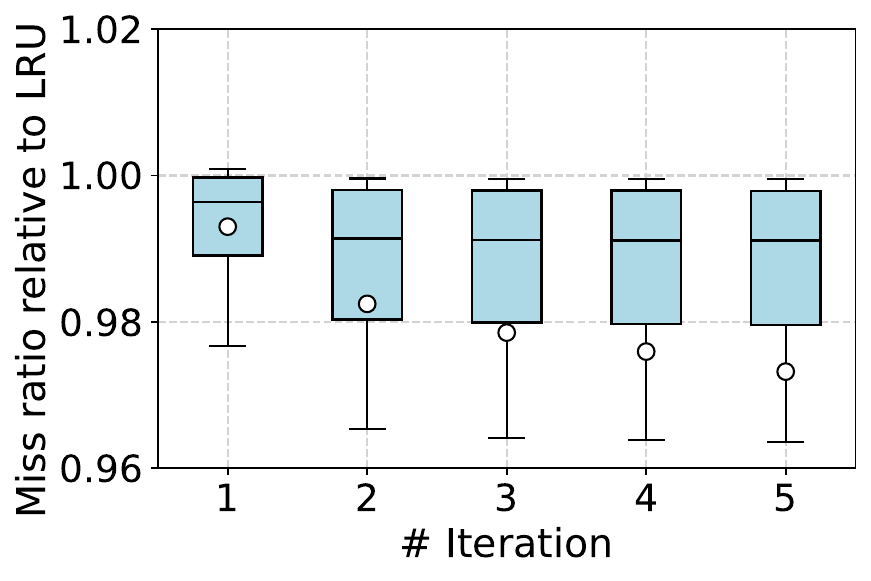}
        \caption{Relative miss ratio}
        \label{fig:bc_miss}
    \end{subfigure}
    \hspace{0.8em}
    \begin{subfigure}[b]{0.28\linewidth}
        \includegraphics[width=\linewidth]{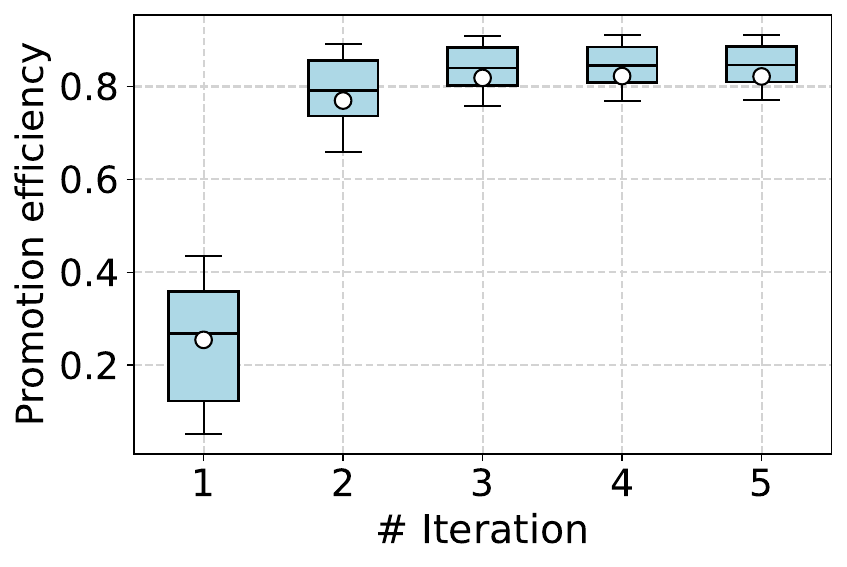}
        \caption{Misses reduced per promotion}
        \label{fig:bc_promotion_efficiency}
    \end{subfigure}
    \vspace{-1.2em}
    \caption{Offline \clock: \normalfont{using future information to filter out unnecessary promotions reduces both promotions and miss ratio. }
}
    \label{fig:clock-belady}
\end{figure*}
\lp techniques have shown that reducing the number of promotions in caching systems can significantly enhance scalability without sacrificing the miss ratio.
In this section, we investigate whether ranking, the goal of promotion, is fundamentally necessary and whether \lp can be even lazier if we have oracle information. More specifically, we use oracle future knowledge as an analytical tool to establish an upper bound on how few promotions are fundamentally needed. 
We first use the future access time obtained from offline analysis to perform early eviction and find that ranking is mostly unnecessary if we can predict whether an object will be requested before it is evicted (\Cref{sec:ranking}). Building on this insight, we find that 
using future information in \clock not only reduces the number of promotions to 6\%, but also reduces the miss ratio by more than 1\% (\Cref{sec:age-insight}), compared to LRU.

\subsection{Object ranking is unnecessary}
\label{sec:ranking}
In this subsection, we evaluate whether ranking objects in the cache is necessary to obtain a low miss ratio. 

\noindent \textbf{Belady early eviction.}
Traditional Belady evicts the object with the longest future reuse distance whenever space is needed. We extend this idea by allowing eviction at \emph{insertion and request time}: if the next use of an object is too far in the future, it can be discarded immediately. We refer to this as \emph{Belady early eviction} (BEE). 
Because BEE only requires a binary decision from the oracle---whether this object will be used before being evicted---it is essentially easier to predict than the exact reuse time. 

\noindent \textbf{Oracle definition.}
Let $R(o)$ denote the reuse distance (number of requests until the next access) of object $o$, and let $S$ denote the average eviction age measured in the number of requests. 
Under Oracle knowledge, early eviction is:
\[
    \text{evict $o$ at access time if } R(o) > S.
\]

\noindent \textbf{Approximate oracle form.}
In practice, the average eviction age $S$ is not directly observable and changes with decisions on other objects. 
We approximate it by
\[
    \widehat S \;\approx\; \frac{\text{cache size}}{\text{miss ratio}},
\]
To tolerate estimation error, we introduce a \emph{tolerance factor}
$\alpha \geq 1$:
\[
    \text{evict $o$ if } R(o) > \alpha \cdot \widehat S.
\]
Here, $\alpha=1$ applies the strict estimator, while $\alpha>1$ retains objects
more conservatively.
When Belady early eviction cannot remove enough objects, we use \random and Random for eviction, denoted as \beladyrandomLRU and \beladyrandom, respectively.

\noindent \textbf{Ranking objects for eviction is not important.}
\Cref{fig:rbl_0.01} and \Cref{fig:rb_0.01} show that BEE significantly reduces miss ratio compared to LRU, achieving a $14\%$ average reduction at tolerance factor~$5$.
Moreover, \beladyrandomLRU and \beladyrandom exhibit similar miss ratios across tolerance factors (except at~$\infty$), indicating that the choice of the fallback eviction algorithm is largely unimportant in the presence of early-eviction signals.
In other words, ranking cached objects for eviction---i.e., promotion---is not important.
In particular, \beladyrandom and \beladyrandomLRU do not rely on promotions; they primarily use future reuse distances to drive most evictions (\Cref{fig:distribution_0.01}) and, by proactively removing cold objects, achieve lower miss ratios than LRU without ranking.


If we employ a binary classifier to predict whether an object will be reused before being evicted. The model can be used to trigger early eviction. 
This differs from existing learned eviction algorithms, e.g., LRB~\cite{song_learning_2020}, which predict each object's reuse distance using regression as an objective~\footnote{Binary classification is simpler than multi-class classification, which is often simpler than regression.}. 


This observation also explains why the new eviction algorithm, such as S3-FIFO~\cite{yang_fifo_2023-1} and SIEVE~\cite{sieve}, achieves state-of-the-art efficiency. 
They evict new objects quickly (called quick demotion) without ranking, but can still achieve a low miss ratio.




\begin{figure*}[h]
    \centering
    \begin{subfigure}[b]{0.28\linewidth}
        \centering
        \includegraphics[width=\linewidth]{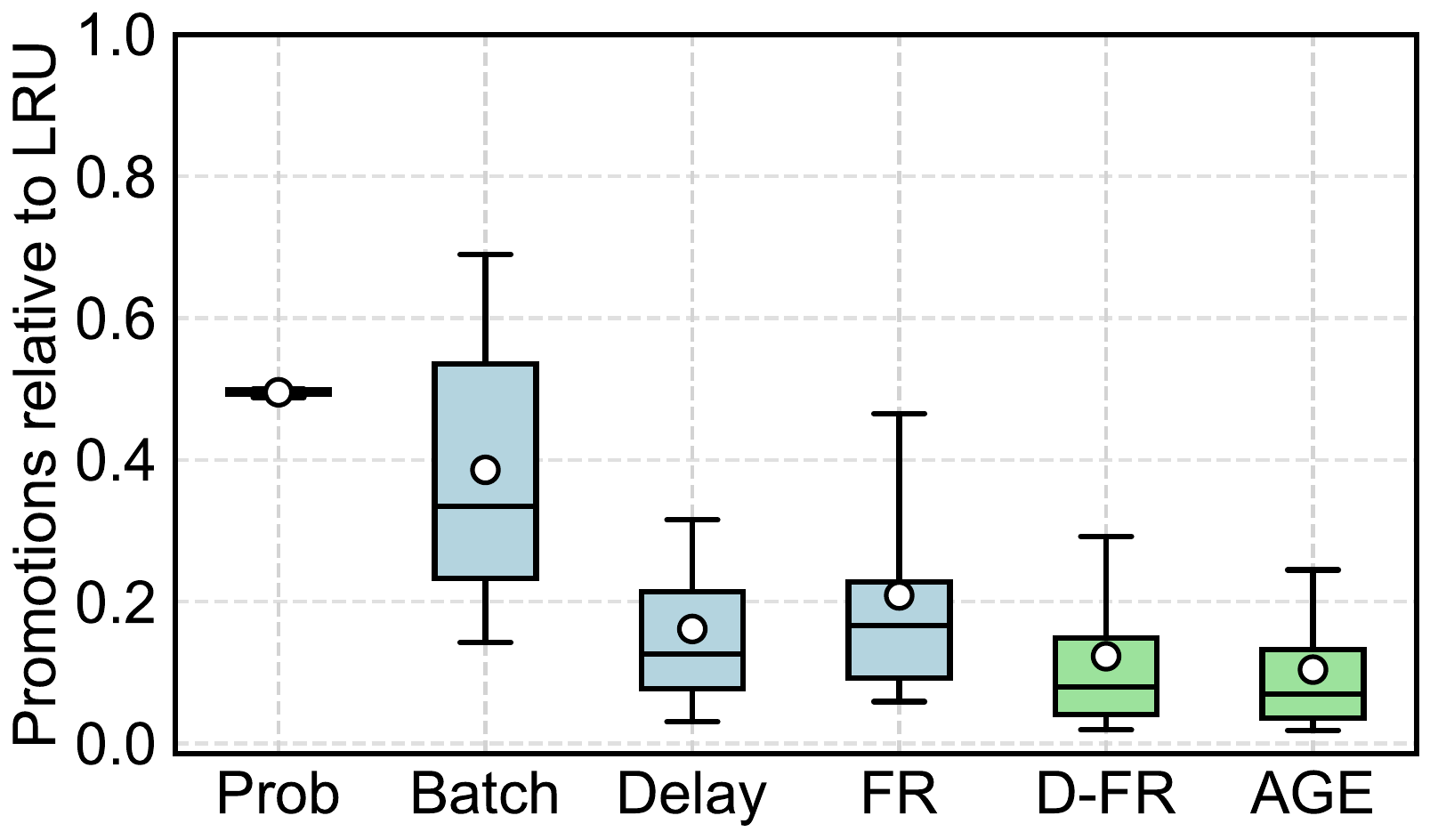}
        \caption{Relative number of promotions}
        \label{fig:prac_promotions}
    \end{subfigure}
    \hspace{0.8em}
    \begin{subfigure}[b]{0.28\linewidth}
        \centering
        \includegraphics[width=\linewidth]{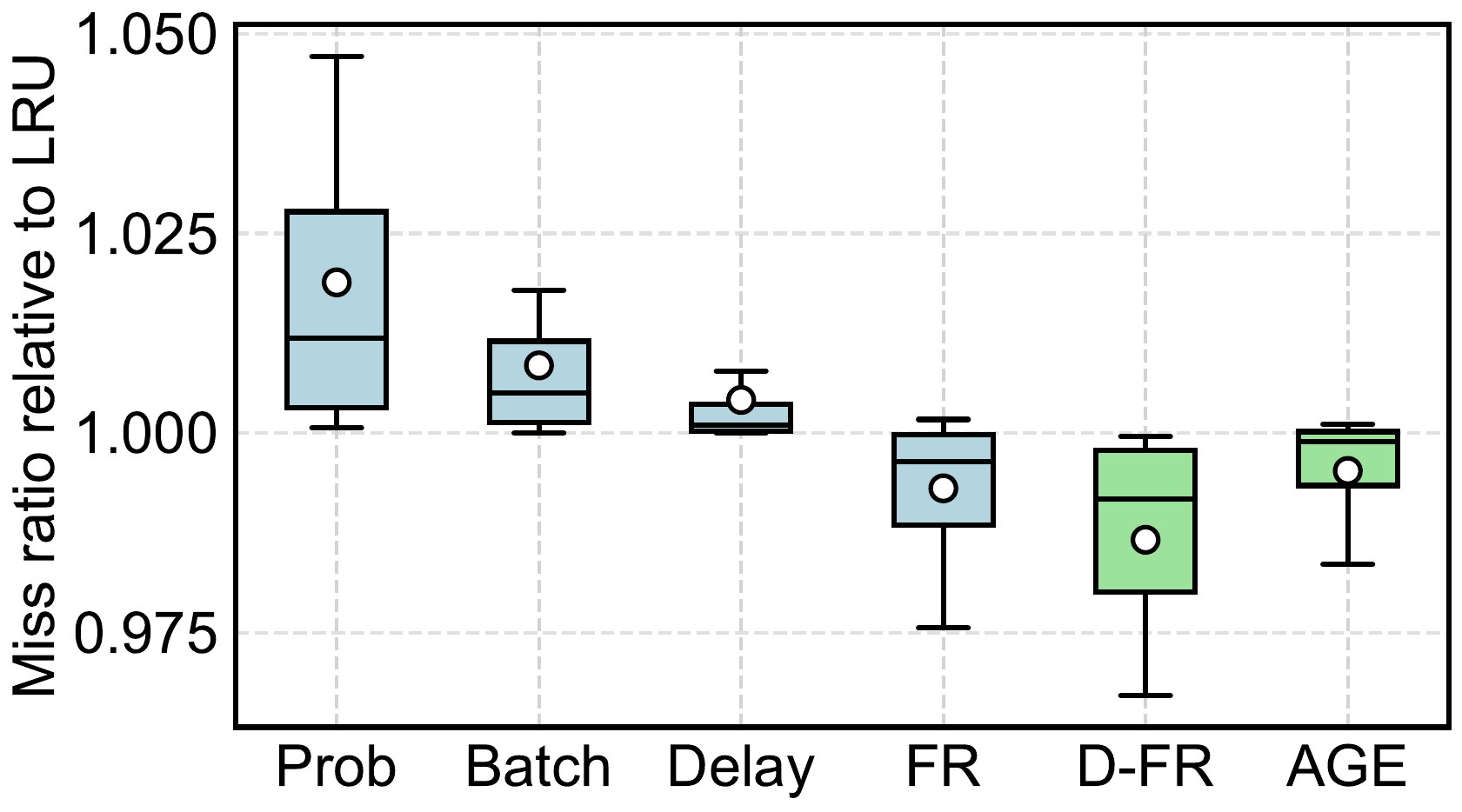}
        \caption{Relative miss ratio}
        \label{fig:prac_miss_ratio}
    \end{subfigure}
    \hspace{0.8em}
    \begin{subfigure}[b]{0.28\linewidth}
        \centering
        \includegraphics[width=\linewidth]{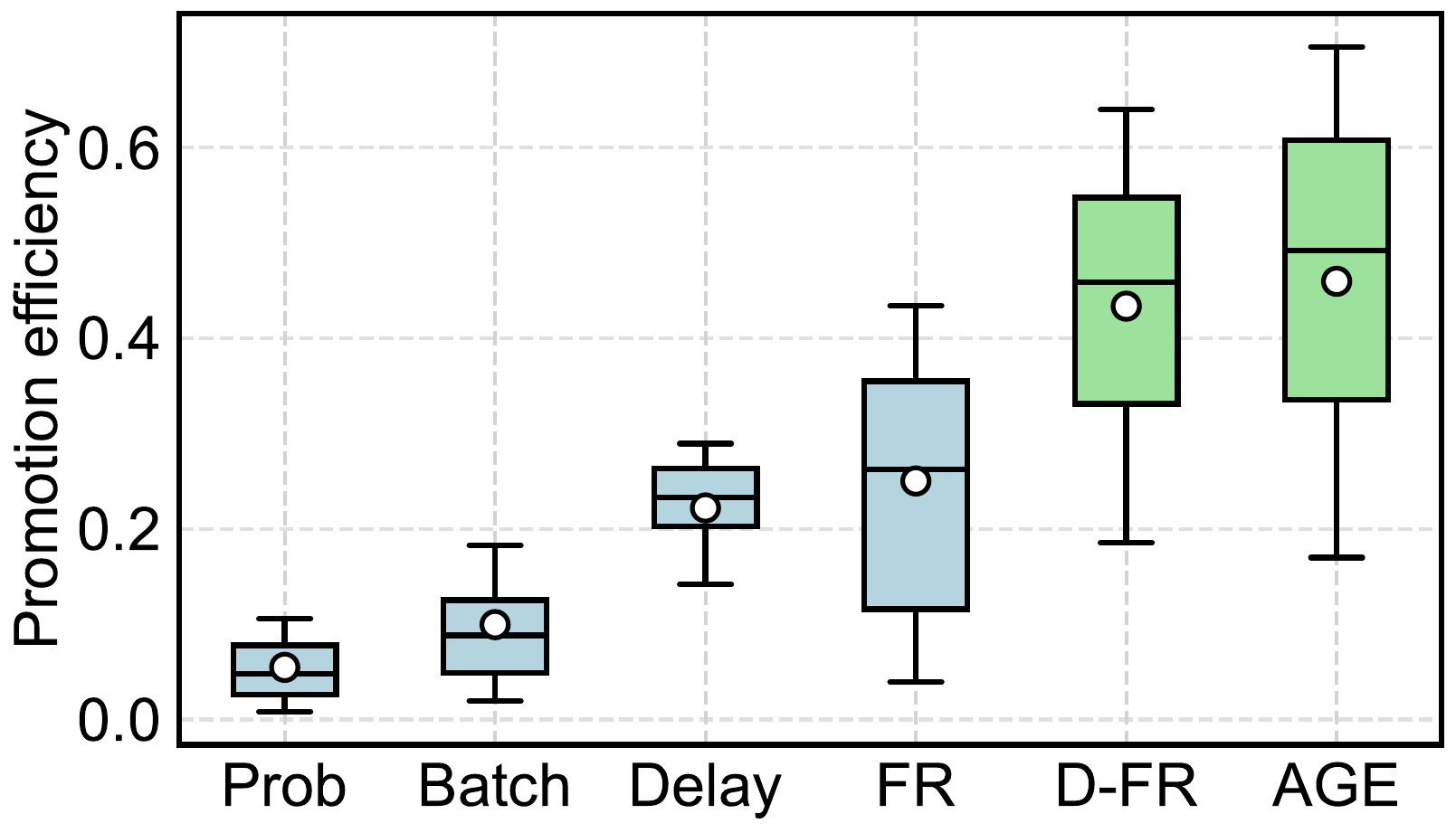}
        \caption{Relative promotion efficiency}
        \label{fig:prac_efficiency}
    \end{subfigure}
    \caption{D-FR and AGE significantly reduce the number of promotions, without compromising the miss ratio. \normalfont{We use the best parameter from each \lp technique that balances between miss ratio and promotions.} }
    \label{fig:prac_result}
\end{figure*}

\subsection{Lazier upon cache misses}
\label{sec:age-insight}
We have demonstrated that most of the promotions in the cache are unnecessary if we have future information. 
However, the previous section assumes that the cache can evict objects at any time and requires a binary prediction to be made at each request, which is computationally expensive. 
In this section, we study how future information can help \clock, which limits promotion and thus binary prediction to eviction time. 
Specifically, we design \beladyfr by not promoting an object if its next access time is too far in the future. 
For any request sequence and cache capacity~$C$, policies that promote objects on hits offer no additional benefit once promotion at eviction is allowed. A policy that only promotes at eviction time is guaranteed to be as good as any policy that promotes at object access time. 
Moreover, promotion decisions at eviction time can leverage more information.


\noindent
\textbf{Offline \clock design.}
Building on FR, we design Offline \clock, which runs multiple iterations through a trace. The zeroth iteration is the same as \clock, however, we mark the promotions (reinsertions) that do not lead to a cache hit. In subsequent iterations, the marked promotions are not performed. Therefore, some promotions are reduced in each iteration. 

\noindent \textbf{Offline FR reduces both promotions and miss ratios.}
\Cref{fig:bc_promotion} shows that \beladyfr can reduce promotions from LRU by more than 90\%, demonstrating the huge potential. 
Reducing the number of promotions in \beladyfr does not bring the side effect of increasing miss ratio. 
We observe that \beladyfr achieves a lower miss ratio compared to \clock (\Cref{fig:bc_miss}), with consistent reductions in both median and mean. As the number of iterations increases, the number of promotions and the miss ratio start to converge. 

\Cref{fig:bc_promotion_efficiency} shows promotion efficiency of \beladyfr. We find that filtering out unnecessary promotions significantly improves promotion efficiency, achieving over 0.8---each promotion reduces 0.8 cache misses. 
This suggests that promotion at eviction is sufficient to achieve both a low number of promotions and a low miss ratio, and the future \lp techniques can be built on top of a FIFO queue using FIFO-reinsertion. 

\section{Practical lazier promotion} \label{sec:modification}
The previous section shows that promotion and ranking objects are not important: if we have future information, we can further reduce the number of promotions without increasing the miss ratio. 
This section demonstrates how to achieve this goal without relying on future information. 

In \Cref{sec:dclock}, we introduce Delayed \clock (D-FR), a practical enhancement to \clock inspired by \delay. \dfr reduces both promotions and miss ratios compared to \clock, however, the promotion reduction is limited. 
In \Cref{sec:age}, we introduce Age-Guided Eviction (AGE). AGE uses recency to predict whether a promotion will be useful during eviction and discards unnecessary promotions. Compared to \dfr, \age can reduce more promotions; however, it increases the miss ratio for some traces compared to \clock. 

 
\subsection{Delayed \clock (D-FR)}
\label{sec:dclock}
\begin{lstlisting}[language=python, style=custom, caption={Delayed FIFO-reinsertion}]
delay_time = delay_ratio * cache_size
# On access 
t = current_time - obj.access_time
if t > delay_time:
    obj.freq += 1    
    obj.access_time = current_time
\end{lstlisting}

\Cref{sec:delay} shows that \delay consistently achieves the best promotion efficiency while keeping the miss ratio close to that of LRU, indicating that many promotions occur too soon after a previous one and bring little value. 
This suggests two principles: (1) do not reward multiple hits in a short window, and (2) if we must reward, do it at the \emph{eviction time}, where the decision is most informed. Building on these, we design \textit{D-FR}, which does not increment the frequency counter upon clustered hits. More specifically, \dfr tracks the last access and insertion time. If the current hit falls within a delay-time threshold, the frequency counter will not increment. 
Similar to \delay, \dfr adds a 4-byte timestamp. The computational overhead from conditional checks (metadata read from CPU cache) is outweighed by the reduction in promotions (metadata write), yielding a 5\% throughput increase over \clock in our evaluations.

\Cref{fig:prac_result} shows that compared to \clock, \dfr further reduces the number of promotions by 60\% for a median trace. Meanwhile, it also reduces the miss ratio similar to offline \clock. As a result, \dfr achieves a higher promotion efficiency compared to existing \lp techniques. 

\begin{figure*}[h]
    \centering
    \begin{subfigure}[b]{0.28\linewidth}
        \centering
        \includegraphics[width=\linewidth]{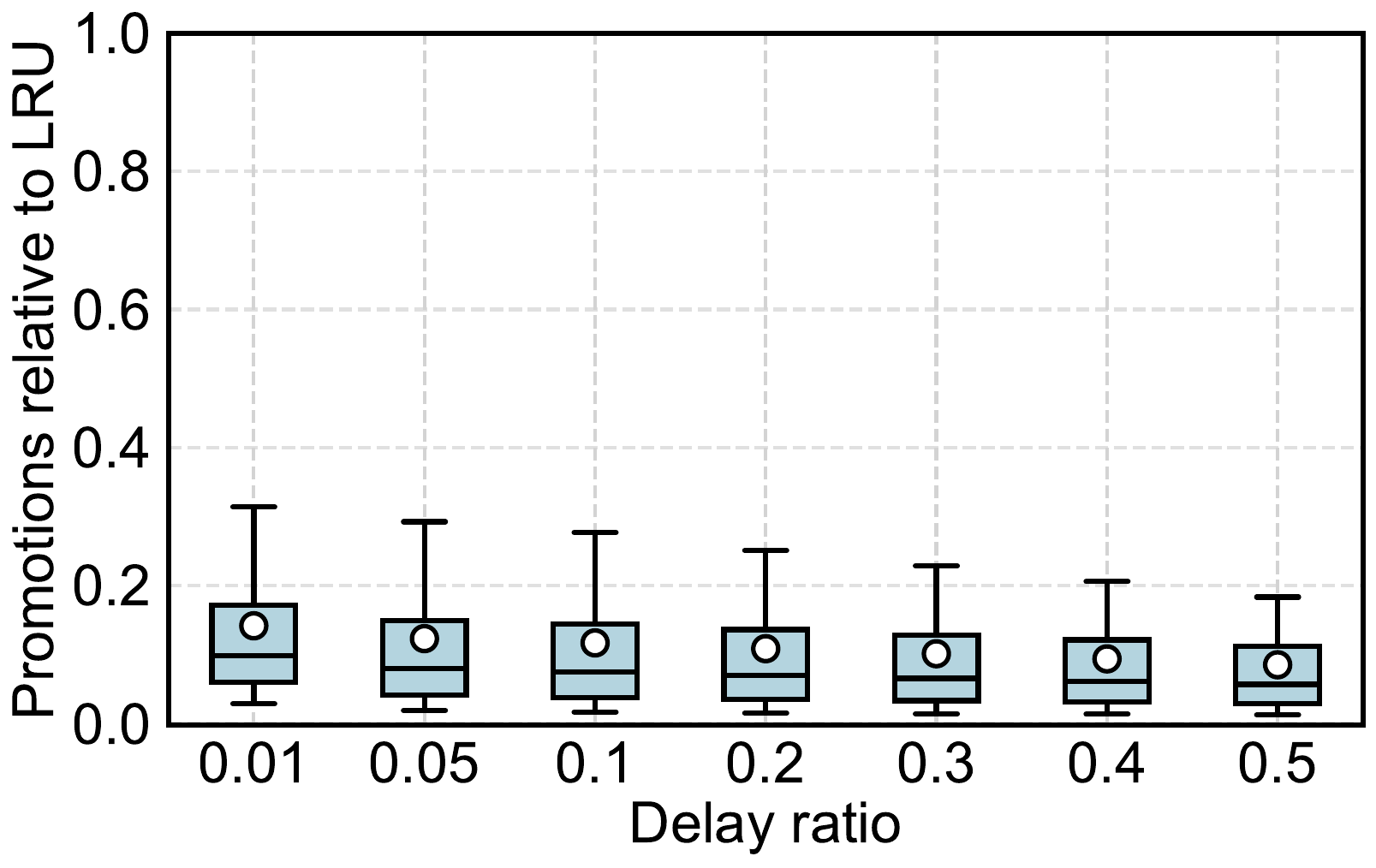}
        \caption{Relative number of promotions}
        \label{fig:ratio_dclock_promotions}
    \end{subfigure}
    \hspace{0.8em}
    \begin{subfigure}[b]{0.28\linewidth}
        \centering
        \includegraphics[width=\linewidth]{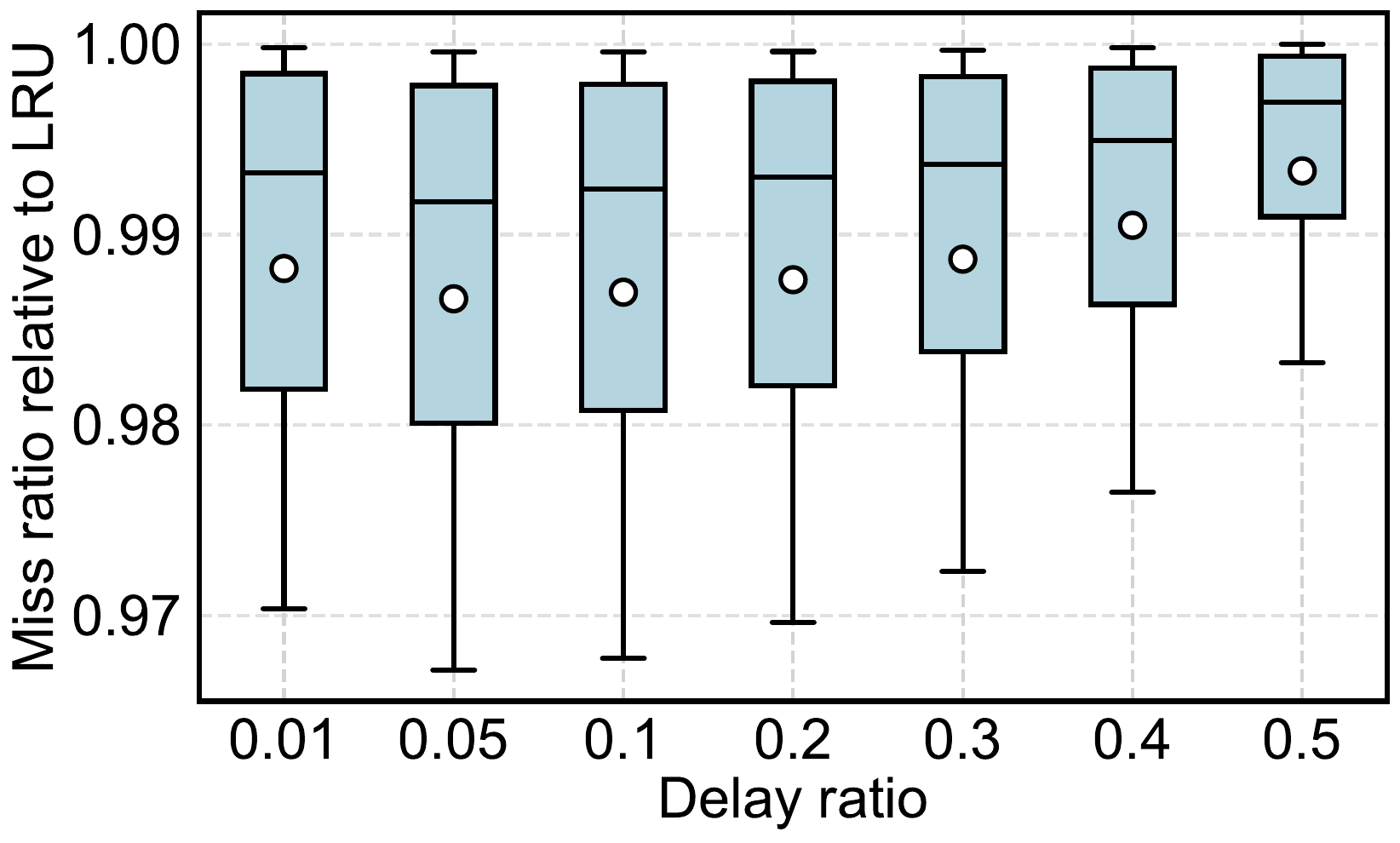}
        \caption{Relative miss ratio}
        \label{fig:ratio_dclock_miss}
    \end{subfigure}
    \hspace{0.8em}
    \begin{subfigure}[b]{0.28\linewidth}
        \centering
        \includegraphics[width=\linewidth]{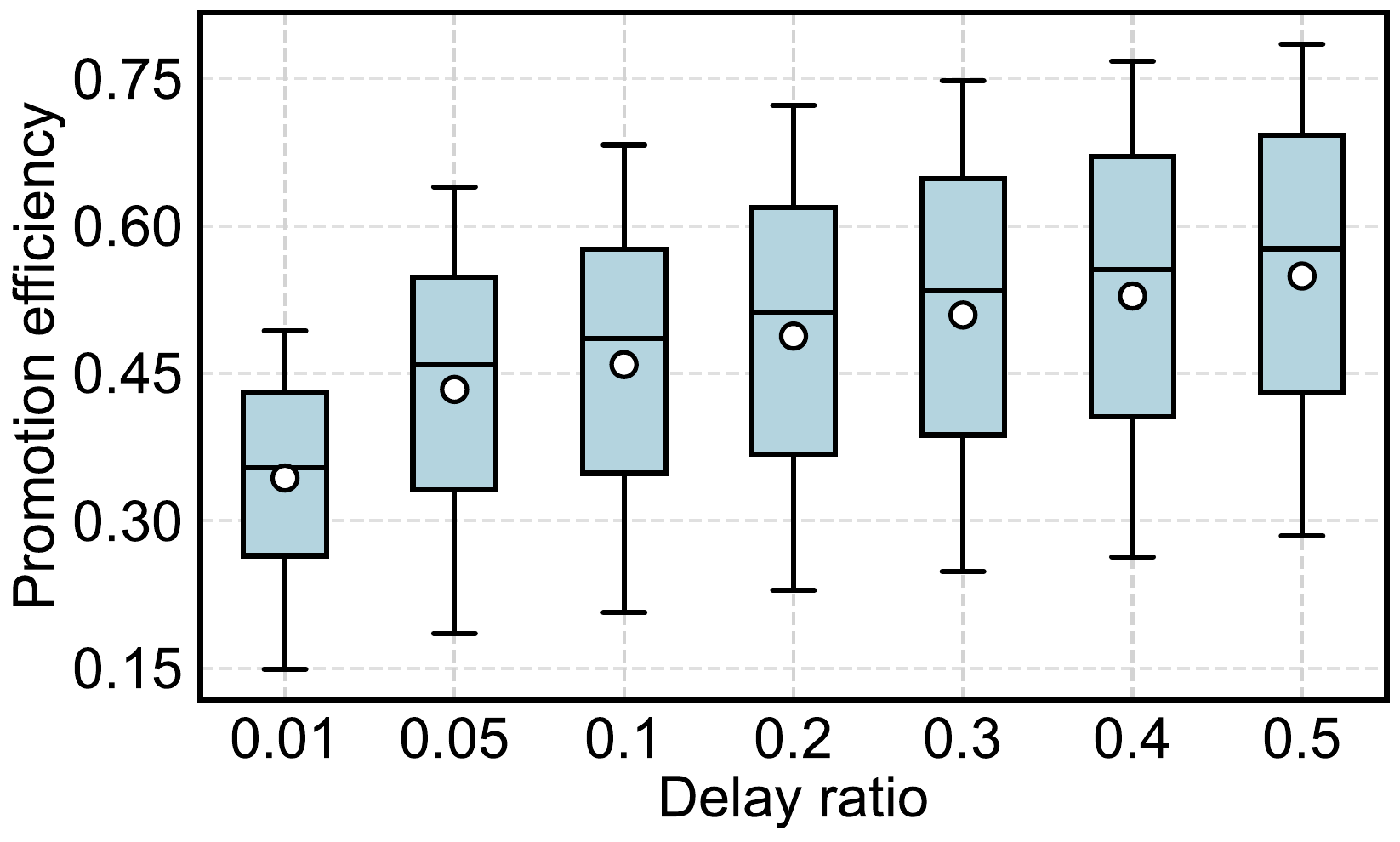}
        \caption{Relative promotion efficiency}
        \label{fig:ratio_dclock_efficiency}
    \end{subfigure}
    \vspace{-1.2em}
    \caption{D-FR is not sensitive to delay ratio. \normalfont{As delay ratio increases, more promotions are reduced at the cost of slightly higher miss ratios.} }
    \label{fig:ratio_dclock_result}
\end{figure*}

The results in~\Cref{fig:prac_result} use a default frequency bit of 1 and a delay ratio of 5\%. We study the sensitivity of delay time in \Cref{fig:ratio_dclock_result}. We find that \dfr is not sensitive to the delay ratio, and different ratios show similar results. Choosing a larger delay ratio allows for a greater promotion reduction, albeit at the cost of slightly higher miss ratios. However, unlike \delay, which increases the miss ratio over LRU, \dfr can consistently achieve a miss ratio lower than LRU.

\begin{figure*}[h]
    \centering
    \begin{subfigure}[b]{0.28\linewidth}
        \centering
        \includegraphics[width=\linewidth]{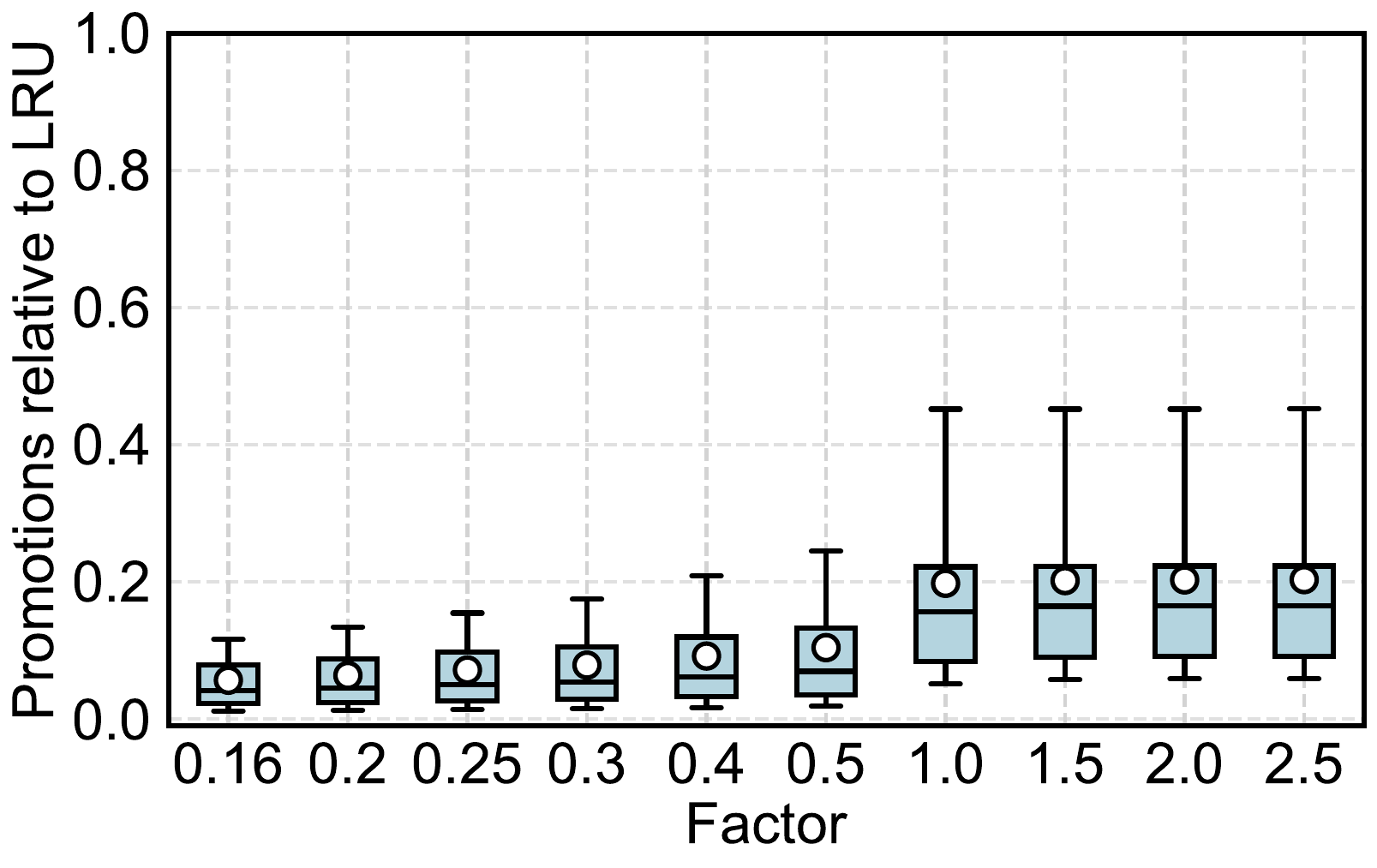}
        \caption{Relative number of promotions}
        \label{fig:predclock_promotions}
    \end{subfigure}
    \hspace{0.8em}
    \begin{subfigure}[b]{0.28\linewidth}
        \centering
        \includegraphics[width=\linewidth]{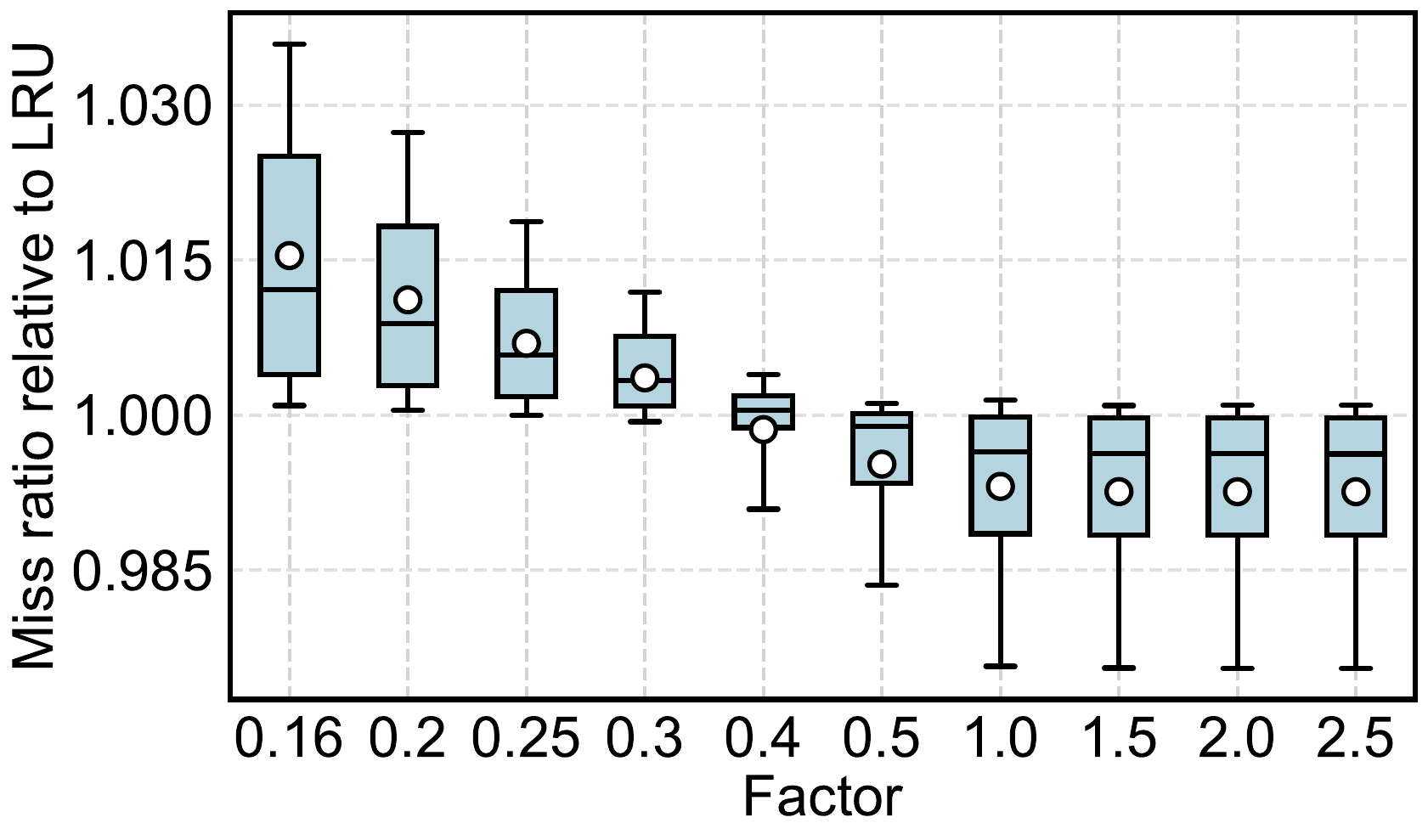}
        \caption{Relative miss ratio}
        \label{fig:predclock_miss_ratio}
    \end{subfigure}
    \hspace{0.8em}
    \begin{subfigure}[b]{0.28\linewidth}
        \centering
        \includegraphics[width=\linewidth]{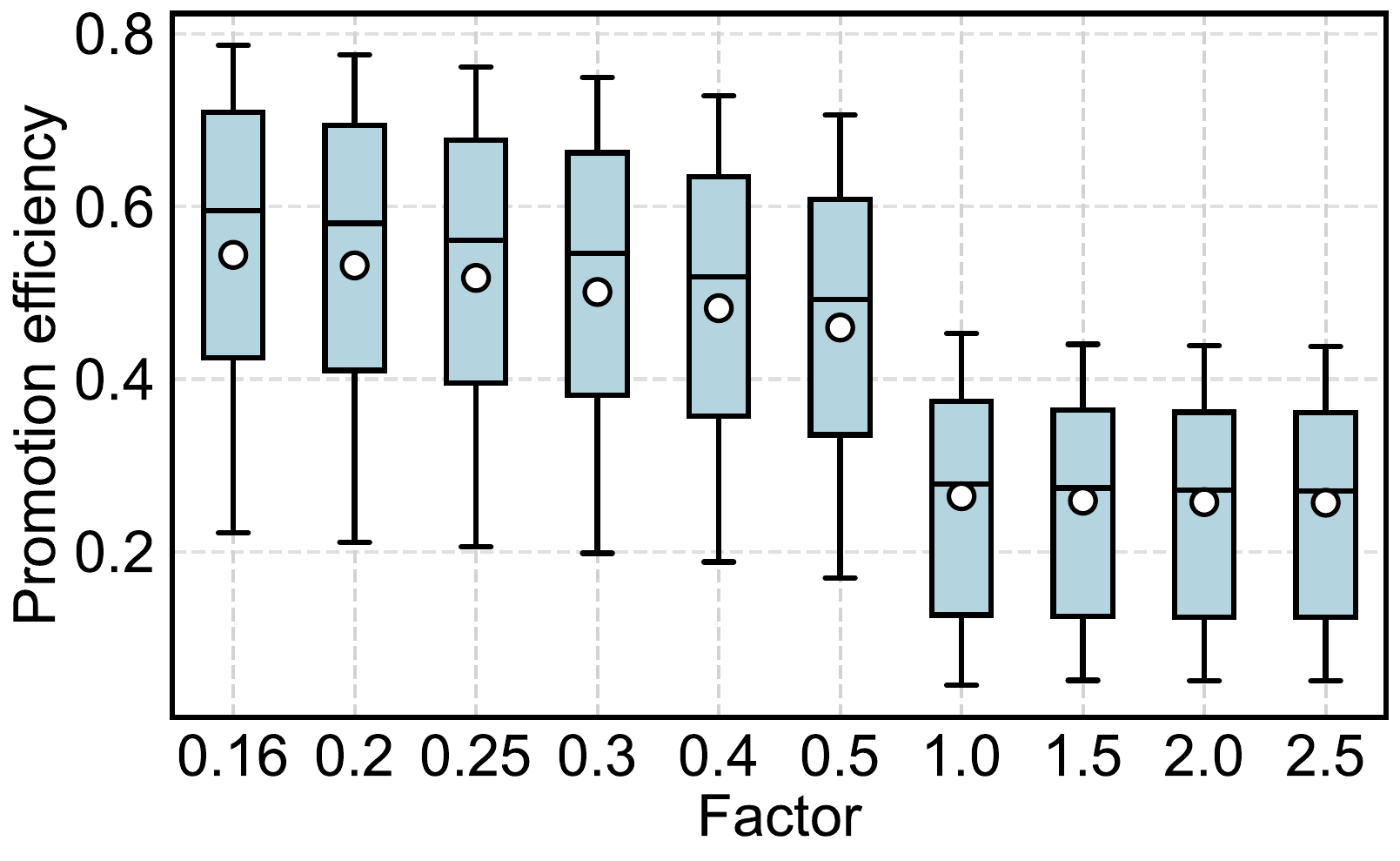}
        \caption{Relative promotion efficiency}
        \label{fig:predclock_efficiency}
    \end{subfigure}
    \vspace{-1.2em}
    \caption{Age is not sensitive to the factor parameter. }
    \label{fig:promotion_predclock}
\end{figure*}

\subsection{Age guided eviction (AGE)}
\label{sec:age}
\begin{lstlisting}[language=Python, style=custom, caption={Age guided eviction (AGE)}, label=algo:age]
# Eviction
threshold = cache_size / miss_ratio * factor
obj_to_evict = queue.front()
while obj_to_evict.freq > 0:
    obj_to_evict.freq -= 1
    t = obj_to_evict.last_access_time
    age = current_time - t
    if (age >= threshold):
        break # evict the object
    else:
        # check the next object
    ...
\end{lstlisting}
\clock makes promotion decisions based on frequency without considering recency information. To further reduce unnecessary promotions in \clock, we design \age, which leverages object age as a filter. 
If the time since the last access to an object has been very long, the object may be unpopular, and the likelihood of it receiving a request will be low. Therefore, \age does not promote objects whose age exceeds a threshold. 
It calculates the threshold similar to BEE (\cref{sec:ranking}) with a tolerance factor (\cref{algo:age}). 

\Cref{fig:prac_result} shows that \age has fewer promotions than \dfr with increased miss ratio. However, unlike \dfr, which reduces miss ratio over \clock, \age achieves a similar or slightly higher miss ratio compared to \clock. As a result, it achieves a similar promotion efficiency to \dfr, around 48\%, and both are significantly higher than \clock, at approximately 24\%. 



We show \age with a tolerance factor of 0.5 in \Cref{fig:prac_result}. To evaluate the sensitivity, we evaluate a wide range of factors in~\Cref{fig:promotion_predclock}. We find that it is straightforward to choose this parameter. A smaller factor reduces promotions more aggressively, while leading to a higher miss ratio; a larger factor has a smaller impact on both promotions and miss ratios. Increasing the factor over a certain threshold results in zero promotions being filtered out, yielding the same algorithm as \clock.

\section{Related Work}
We have covered the most relevant works in \Cref{sec:bg}. 
This section provides additional works related to caching and workload measurements. 

\noindent \textbf{Cache efficiency and scalability. }
The concept of \lp is first introduced in a HotOS work~\cite{yang_fifo_2023_manual}, where the authors show that eviction algorithms should use lazy promotion to improve throughput and scalability, and quick demotion to improve efficiency. However, it does not delve into different lazy promotion techniques. 
Many other works also improve a cache's efficiency by designing a better eviction algorithm, such as ARC~\cite{megiddo_arc_2003}, LIRS~\cite{jiang_lirs_2002, li_dlirs_2018, zhong_lirs2_2021}, TinyLFU~\cite{einziger_tinylfu_2017,zhang_tinyllama_2024}, 2Q~\cite{johnson_2q_1994}, LRFU~\cite{donghee_lee_lrfu_2001}, GDSF~\cite{cherkasova_improving_1998}, LeCaR~\cite{vietri_driving_2018}, CACHEUS~\cite{rodriguez_learning_2021}, LHD~\cite{beckmann_lhd_2018}, LRB~\cite{song_learning_2020}, HALP~\cite{song2023halp}, GL-Cache~\cite{yang_gl-cache_2023_manual}, SIEVE~\cite{sieve}. Most of these algorithms typically have lower throughput compared to LRU due to computational overhead. There are also many works that improve the throughput and scalability of caches, such as MemC3~\cite{fan_memc3_2013}, Segcache~\cite{yang_segcache_2021}, and BP-wrapper~\cite{ding_bp-wrapper_2009}.
Unlike traditional system designs, this work focuses on understanding different techniques that have already been deployed in production. 

\noindent\textbf{Workloads and performance measurement.} 
This work uses 6357 traces to characterize different \lp techniques. Many previous works have characterized production systems or studied the workloads of production systems. 
Juncheng performed a detailed study of over 100 Memcached cache clusters at Twitter~\cite{yang_large_2020}; Siying conducted a comprehensive study of the RocksDB deployments at Meta~\cite{dong2021rocksdb}; Nishtala and Atikoglu studied the Memcached deployment challenges and workloads at Meta~\cite{nishtala_scaling_2013, atikoglu_workload_2012}. Additionally, Alibaba and Tencent have conducted detailed workload studies on their elastic block storage cloud offerings~\cite {alibaba-trace1, tencent_block}.
Moreover, while many studies on eviction-algorithm design evaluate multiple algorithms~\cite{kirilin_rl-cache_2019,song_learning_2020, blankstein_hyperbolic_2017, sieve, hu_raven_2022}, they rarely examine in depth those not proposed in the paper. 
Ziyue shows that aggressively increasing hit ratio can hurt cache throughput due to contention on promotion, but our work demonstrates that suppressing low-value promotions can simultaneously preserve miss ratio and significantly improve scalability~\cite{can_increasing}.
To the best of our knowledge, this work is the first to examine the effectiveness of the widely deployed lazy promotion techniques. 

\section{Conclusion}
Although \lp techniques are widely used in production systems, there has been very limited understanding about their effectiveness. 
This paper presents a comprehensive evaluation of \lp techniques in cache evictions. 
We find that \delay and \clock are effective at reducing promotions while maintaining miss ratios. However, promotion reductions in \prob come at a cost of increased miss ratios, and \batch is heavily workload-dependent. 
In addition, we identify that most cache promotions are unnecessary, exposing considerable opportunities for optimization. We introduce two new simple methods, \dfr and \age, and demonstrate that they can effectively reduce the number of promotions without compromising miss ratios. 


\begin{acks}
We thank the anonymous reviewers for their constructive feedback and suggestions. 
\end{acks}


\clearpage
\bibliographystyle{ACM-Reference-Format}
\bibliography{references_jason, references_other, misc, manual}

\end{document}